\documentclass[AMS,Times1COL]{WileyNJDv5} 

\usepackage[markup=underlined]{changes}
\usepackage{cancel}
\definechangesauthor[name={Main Author}, color=red]{MA}

\articletype{Original Research Paper}%

\received{Date Month Year}
\revised{Date Month Year}
\accepted{Date Month Year}
\journal{Journal}
\volume{00}
\copyyear{2023}
\startpage{1}

\newcommand{\cor}{\color{black}} 
\newcommand{\fin}{\color{black}} 

\begin{document}

\title{Exploring Polarimetric properties preservation for PolSAR image reconstruction with Complex-valued Convolutional Neural Networks}

\author[1,2]{Quentin Gabot}

\author[2]{Joana Frontera-Pons}

\author[1,3]{Jérémy Fix}

\author[1]{Chengfang Ren}

\author[1,2]{Jean-Philippe Ovarlez}

\authormark{GABOT \textsc{et al.}}
\titlemark{Exploring Polarimetric Properties Preservation during Reconstruction of PolSAR images using Complex-valued Convolutional Neural Networks}

\address[1]{\orgdiv{Université Paris‐Saclay, CentraleSupélec}, \orgname{SONDRA}, \orgaddress{91190 Gif-sur-Yvette, \country{France}}}

\address[2]{\orgdiv{DEMR}, \orgname{ONERA}, \orgaddress{91120 Palaiseau, \country{France}}}

\address[3]{Université de Lorraine, CentraleSupélec, CNRS,
LORIA, Metz, \country{France}}

\corres{Quentin Gabot, Université Paris‐Saclay, CentraleSupélec \& ONERA, 91190 Gif‐sur‐Yvette, France. \email{quentin.gabot@centralesupelec.fr}}


\abstract[Abstract]{The inherently complex-valued nature of Polarimetric SAR data necessitates using specialized algorithms capable of directly processing complex-valued representations. However, this aspect remains underexplored in the deep learning community, with many studies opting to convert complex signals into the real domain before applying conventional real-valued models. In this work, we leverage complex-valued neural networks and investigate the performance of complex-valued \cor Convolutional AutoEncoders\fin. We show that these networks can effectively compress and reconstruct fully polarimetric SAR data while preserving essential physical characteristics, as demonstrated through Pauli, Krogager, and Cameron coherent decompositions, as well as the non-coherent $H-\alpha$ decomposition. Finally, we highlight the advantages of complex-valued neural networks over their real-valued counterparts. These insights pave the way for developing robust, physics-informed, complex-valued generative models for SAR data processing.}

\keywords{Complex-valued Auto-Encoders, PolSAR image reconstruction, polarimetric decompositions}


\maketitle

\renewcommand\thefootnote{}
\footnotetext{\textbf{Abbreviations:} SAR, Synthetic Aperture Radar; PolSAR, Polarimetric Synthetic Aperture Radar; AE, AutoEncoder; CVNN, complex-valued neural network; RVNN, real-valued neural network.}

\renewcommand\thefootnote{\fnsymbol{footnote}}
\setcounter{footnote}{1}

\section{Introduction}
\label{sec:intro}

%
Synthetic Aperture Radar (SAR) polarimetry is a well-established technique for extracting qualitative and quantitative physical information across various domains, including land, snow and ice, ocean, and urban environments. It relies on analyzing the polarimetric characteristics of natural and artificial scatterers \cite{henderson1998manual}. Like other SAR imaging techniques, such as interferometry and tomography, SAR polarimetry offers high-resolution, all-weather, day-and-night imaging capabilities, making it invaluable for applications in Geoscience, climate change studies, and environmental and Earth system monitoring \cite{henderson1997sar}. As a result, Polarimetric SAR (PolSAR) imaging significantly departs from the conventional optical imaging methods commonly used in the computer vision field. Due to challenges such as speckle and the inherently complex-valued nature of PolSAR data, standard computer vision techniques are not directly applicable. \\
To address this limitation, the scientific community has frequently discarded phase information, retaining only amplitude data to convert SAR images into single-channel, real-valued representations. This transformation enables the application of well-established real-valued neural networks \cite{frontera2023unsupervised}, which are inherently more suitable for traditional optical imagery than for the SAR data. However, phase information is essential for advanced SAR techniques such as polarimetry and interferometry. It should not be omitted by machine learning algorithms aiming to fully exploit the potential of SAR data, such as polarimetry and interferometry, as it carries the information of interest to extract polarimetric properties or build digital elevation models.
\\
\\ 
{\cor Many classical feature extraction methods have been explored in the SAR imaging literature, such as Principal Component Analysis (PCA) \cite{azimi2002terrain}, Morphological Component Analysis (MCA) \cite{dominguez2015fully}, and Dictionary Learning \cite{zhan2013sar}. However, all these techniques rely on linear combinations of the learned features, which can hinder accurate representation of the highly nonlinear and complex structures present in SAR images.}
In this paper, we advance the study of complex-valued Convolutional AutoEncoders applied to SAR data by demonstrating their effectiveness in preserving polarimetric properties during the reconstruction of fully polarimetric SAR images. The reconstruction task can also be considered a foundational step toward data generation. Notably, both Autoencoders (AEs) \cite{hinton1993autoencoders} and Variational Autoencoders (VAEs) \cite{kingma_auto-encoding_2013} share the core mechanism of compressing input images into latent representations. While traditional AEs do not explicitly model the underlying data distribution, VAEs can generate synthetic data by sampling from a learned latent distribution. Although generative modeling is not the focus of this study, we demonstrate its feasibility by validating that polarimetric information can be reliably preserved through the reconstruction process. This process mirrors the decoding step in generative models.\\
This work is an extension of our previous study \cite{gabotpreserving}, validating our approach on additional polarimetric decompositions, such as the Cameron decomposition. Extensive experimental results are provided comparing complex-valued AutoEncoders against their real-valued counterparts. \\

\textbf{The key contributions of this work are:}
\begin{itemize}
\item The demonstration of the preservation of various polarimetric properties using complex-valued Convolutional AutoEncoders,
\item An extensive study on the impact of the size of the latent vector on the preservation of polarimetric properties,
\item A comparison of the performance of real-valued and complex-valued Convolutional AutoEncoders regarding the preservation task.
\end{itemize}
In Section \ref{sec:related}, we review the existing works related to polarimetric synthetic aperture radar and complex-valued neural networks. In Section \ref{sec:polsar_ae}, we present the polarimetric properties studied in this article, as well as the theory behind complex-valued \cor Convolutional \fin AutoEncoders. Section \ref{sec:experiments} describes an extensive set of experiments to investigate how well the complex-valued Convolutional AutoEncoders preserve the polarimetric properties. Finally, Section \ref{sec:discussions} highlights some challenges this method poses, some of which are opportunities for grounded applications in remote sensing \cite{mian2019design}, medical imaging \cite{dedmari2018complex}, and Earth monitoring \cite{patruno2013polarimetric}.  \\
\\
\textbf{Notations.} Italic type indicates a scalar quantity $x$, lower case boldface indicates a vector quantity $\mathbf v$, and upper case boldface indicates a matrix or tensor $\mathbf A$. $\Re{(z)}$ and $\Im{(z)}$ respectively denote the real and imaginary parts of complex number $z\in \mathbb{C}$. $\lfloor x \rfloor$ is the mathematical floor function. The modulo operation is denoted as $\mathrm{mod}$. For any vector $\mathbf{v}$, $\mathbf{v}^T$ denotes the transpose operator and $\mathbf{v}[.]$ is the indexing operator. $\left\|\mathbf{v}\right\|_2$ stands for the Euclidean norm, and $\mathrm{vec}(.)$ is the operator that transforms a matrix into a vector by reshaping its elements. The operator $\odot$ is the Hadamard element-wise product between vectors. The composition operator $\circ$ is defined by $(f \circ g)(.)=f(g(.))$. Finally, $\mathcal{N}$, $\mathcal{CN}$, and $\mathcal{U}$ denote the Normal distribution, the complex Normal distribution, and the Uniform distribution. 

\section{Related Works}
\label{sec:related}
\textbf{Polarimetric Synthetic Aperture Radar Imaging.} Synthetic Aperture Radar (SAR) \cite{Moreira} is an advanced form of radar technology. Radar systems detect, track, and identify objects using radio waves, which are situated between microwaves and infrared waves within the electromagnetic spectrum. Electromagnetic waves are oscillations of electric and magnetic energy that propagate in a specific polarization state through space. One can leverage the change in polarization between the emission and reception of the wave to characterize the electromagnetic nature of the scatterer; this is the founding principle of Polarimetric Synthetic Aperture Radar. \\
While PolSAR images are naturally rich in physical information, decompositions are needed to characterize how the target reflects the radar signal in different polarization configurations. Coherent decompositions assume that each resolution cell represents a single, predominant scattering mechanism that a simple geometric structure can model. Non-coherent decompositions postulate that each resolution cell lacks a
dominant scattering mechanism and that the extraction of information requires second-order statistics. Coherent decompositions are particularly suited for studying coherent scatterers, i.e., manufactured objects (vehicles, buildings, etc.). On the other hand, non-coherent decompositions are designed to study natural scatterers (vegetation, mountains, etc.).
\\
\\
One of the most prominent coherent decompositions is the Pauli decomposition \cite{alberga2004potential}, based upon the Pauli matrices \cite{pauli1941relativistic}. To produce physically meaningful results, Krogager et al. proposed another decomposition, based on the response of a sphere, a dihedral, and a helix \cite{Krogager92, Krogager95, Krogager95A}. Finally, Cameron et al. introduced a new decomposition that relies on the properties of two scatterers: reciprocity and symmetry \cite{cameron1990feature, cameron2002simulated}.
\\
\\
Regarding non-coherent decompositions, we can also distinguish several methods. Pottier et al. introduced the $H-\alpha$ decomposition, based upon the decomposition in eigenvalues and eigenvectors of the coherency matrix, computed for each pixel in the image \cite{cloude1996review, cloude2002entropy}. Additionally, Freeman et al. proposed a decomposition that expresses the coherency matrix into multiple components corresponding to various
physical scattering mechanisms \cite{freeman2002three}. Finally, Yamaguchi et al. introduced an extension of the Freeman decomposition to take into account non-reflection symmetric cases \cite{yamaguchi2005four}. In this article, we will study how the properties of some of these decompositions, namely the Pauli, Krogager, Cameron, and  $H-\alpha$ decompositions, are preserved during the reconstruction process of a complex-valued Convolutional AutoEncoder.
\\
\\
\textbf{Complex-valued neural networks.} Complex-Valued Neural Networks (CVNNs) are deep neural networks whose operations and inputs are defined in the complex domain \cite{trabelsi2017deep, Barrachina2021}. Despite the novelty of complex-valued neural network theory, several critical questions regarding CVNNs are still open, one being the optimal representation of their complex nature. \\
Based on the isomorphism between $\mathbb{C}$ and $\mathbb{R}^2$, CVNNs can be either represented as~\cite{lee2022complex}:
\begin{itemize}
    \item A dual-RVNN is a neural network with representations in $\mathbb{R}$, where real and imaginary parts are stacked at the input. Operations are then performed in $\mathbb{R}$,
    \item A split-CVNN processes representations in $\mathbb{R}^2$ with operations in $\mathbb{R}$. Compared to a dual-RVNN, we keep the relationship between real and imaginary components, although the weights are also real-valued.
    \item A full-CVNN contains both representations and operations in $\mathbb{C}$.
\end{itemize} 
Finally, Wu et al. \cite{wu2024complex} recently proved that complex-valued neurons can learn more than real-valued neurons, reigniting discussions surrounding this question. This question goes beyond complex-valued neural networks and is also fundamental for the broader class of vector-valued neural networks~\cite{Valle2024}. \\
In addition to theoretical advancements, standard real-valued neural network (RVNN) architectures need to be extended to the complex domain. Most of the essential elements of neural networks have already been extended to the complex domain. For instance, Li et al. \cite{li2008complex} proposed using Wirtinger calculus to circumvent some issues with complex-valued backpropagation. More recently, Trabelsi et al. \cite{trabelsi2017deep} extended the batch normalization layer to the complex domain. Eilers et al. \cite{Eilers2023} introduced the complex-valued attentional head, a fundamental building block of the cutting-edge transformers. \\
One of the most commonly used complex-valued architectures is the complex-valued convolutional neural network, as it enables the processing of complex-valued imaging modalities, such as synthetic aperture radar (SAR) imaging \cite{barrachina2023comparison, gabotpreserving, zhu2021deep} and magnetic resonance imaging (MRI) \cite{cole2021analysis}. We acknowledge that complex-valued linear AutoEncoders have been introduced by Baldi et al. \cite{baldi2012complex}. More recently, complex-valued \cor Convolutional \fin AutoEncoders have been discussed by Shang et al. \cite{shang_complex-valued_2019}. Interestingly, complex-valued Convolutional AutoEncoders have been extensively studied for object discovery, firstly by Lowe et al. \cite{lowe_complex-valued_2022}. We note that complex-valued Convolutional AutoEncoders have already been tested on dual-polarization SAR imaging by Asiyabi et al. in \cite{asiyabi2022complex}.
\\
\\
\cor While recent studies have explored complex-valued AutoEncoders for tasks such as object discovery or dual-polarization reconstruction, these works primarily assess performance through signal-fidelity metrics (e.g., MSE, coherence preservation). Our work advances this domain by systematically evaluating the preservation of high-level polarimetric semantics. Instead of focusing solely on reconstruction errors, we investigate whether the complex-valued latent representations maintain the physical scattering mechanisms characterized by Pauli, Cameron, and $H-\alpha$ decompositions, which are critical for interpretable remote sensing applications. Our former work introduced the preliminary results of our approach in \cite{gabotpreserving}. In this paper, we significantly extend that work by providing additional theoretical foundations for polarimetric decompositions and CVNNs. We also extend the set of experiments by adding a comparison with dual-RVNNs. \fin
\section{Reconstruction of Polarimetric Properties using Complex-valued AutoEncoders}
\label{sec:polsar_ae}
\subsection{Polarimetric Decomposition}
\label{sec:polsar}
SAR polarimetry utilizes the polarization properties of electromagnetic waves to determine the polarimetric characteristics of scatterers. This technique provides qualitative and quantitative physical information related to ground physics (buildings, vehicles, forests, seas, roads, crops, etc.). Thus, each pixel of the image is associated with a complex scattering matrix $S$, called the Sinclair matrix:
 \begin{equation}
 \mathbf{S} = \begin{pmatrix}
 S_{hh} & S_{vh} \\
 S_{hv} & S_{vv} \\
 \end{pmatrix}
 \end{equation}
The indices $h$ and $v$ refer to the horizontal and vertical polarization states, but any other orthogonal polarimetry basis could be used.
\\
\\
\textbf{Pauli decomposition.} 
The Pauli decomposition expresses the Sinclair matrix $\mathbf{S}$ in the Pauli basis \cite{cloude1996review, lee2017polarimetric}. In the conventional linear, horizontal, and vertical polarization basis, the Pauli basis consists of the following four matrices:
\begin{equation}
     \mathbf{S} =  \alpha\,\mathbf{S}_a+\beta\,\mathbf{S}_b+\gamma\,\mathbf{S}_c + \delta \, \mathbf{S}_d\, ,
\end{equation}
where
\begin{equation}
  \mathbf{S}_a=\frac{1}{\sqrt{2}}\begin{pmatrix} 
   1 & 0 \\ 0 & 1 \end{pmatrix}\,,  \hspace{1cm}
  \mathbf{S}_b=\frac{1}{\sqrt{2}}\begin{pmatrix} 1 & 0\\ 0 & -1, \end{pmatrix}\,, \hspace{1cm}
  \mathbf{S}_c=\frac{1}{\sqrt{2}}\begin{pmatrix} 0 & 1\\ 1 & 0 \end{pmatrix}\,, \hspace{1cm}
  \mathbf{S}_d=\frac{1}{\sqrt{2}}\begin{pmatrix} 0 & -1\\ 1 & 0 \end{pmatrix}\,,
\end{equation}
In the monostatic configuration where the reciprocity principle can be applied \cite{quegan_understanding_2004}, $S_{hv} = S_{vh}$. The Pauli basis then reduces to the matrix family $\mathbf{S}_a$, $\mathbf{S}_b$, and $\mathbf{S}_c$. Consequently, a Sinclair matrix $\mathbf{S}$ can be decomposed as follows:
\begin{equation}
     \mathbf{S} =  \alpha\,\mathbf{S}_a+\beta\,\mathbf{S}_b+\gamma\,\mathbf{S}_c \, .
     \label{eq:lexico}
\end{equation}
The decomposition coefficients are given by $\alpha = \left(S_{hh} + S_{vv}\right)/\sqrt{2}$, $\beta = \left(S_{hh} - S_{vv}\right)/\sqrt{2}$ and $\gamma = \sqrt{2} \, S_{hv}$. They represent the part of the response of a plate observed at normal incidence or a sphere ($\mathbf{S}_a$), the characteristic of a horizontal metallic dihedral ($\mathbf{S}_b$), and the scattering matrix of a metallic dihedral oriented at $45^\circ$ concerning the radar line of sight ($\mathbf{S}_c$), respectively. 
\\
\\
This decomposition defines the Pauli vector as $\displaystyle \mathbf{k} = \left( \alpha, \beta, \gamma \right)^T$. The Pauli vector can be assimilated to an RGB image by taking the modulus of each component. Furthermore, the Pauli vector can also be used to estimate the coherence matrix in Eq. (\ref{eq:coherence}).
\\
\\
\textbf{Krogager decomposition.} A refined alternative approach, proposed by \cite{Krogager92,KrogagerTGRS1993, Krogager95, Krogager95A}, considers a scattering matrix as the combination of the responses of a sphere, an oriented diplane, and a helix:
\begin{equation}
  \mathbf{S}=\exp{(j \varphi)}\, \left(\exp{(j \varphi_s)}\,k_s\,\mathbf{S}_{s}+k_d\,\mathbf{S}_{d}
    (\vartheta)+k_h\,\mathbf{S}_{h} (\vartheta)\right)\, ,
  \label{eq:krog_1}
\end{equation}
where $\mathbf{S}_{s}=\begin{pmatrix}  1 & 0\\0 &
      1\end{pmatrix}$,  $\mathbf{S}_{d} (\vartheta)=\begin{pmatrix}
      \cos 2\vartheta & \sin 2\vartheta\\\sin 2\vartheta &
      -\cos 2\vartheta \end{pmatrix}$ and $\mathbf{S}_{h} (\vartheta)=\exp{(\pm 2 j
    \vartheta)}\, \begin{pmatrix}  1 & \pm j\\ \pm j &
      -1\end{pmatrix}$, where the $\pm$ sign in the helix component varies the left or right-handedness, and it has to be fixed during the estimation of its components.

\begin{figure}[htbp]
    \centering
    \includegraphics[width=0.66\columnwidth]{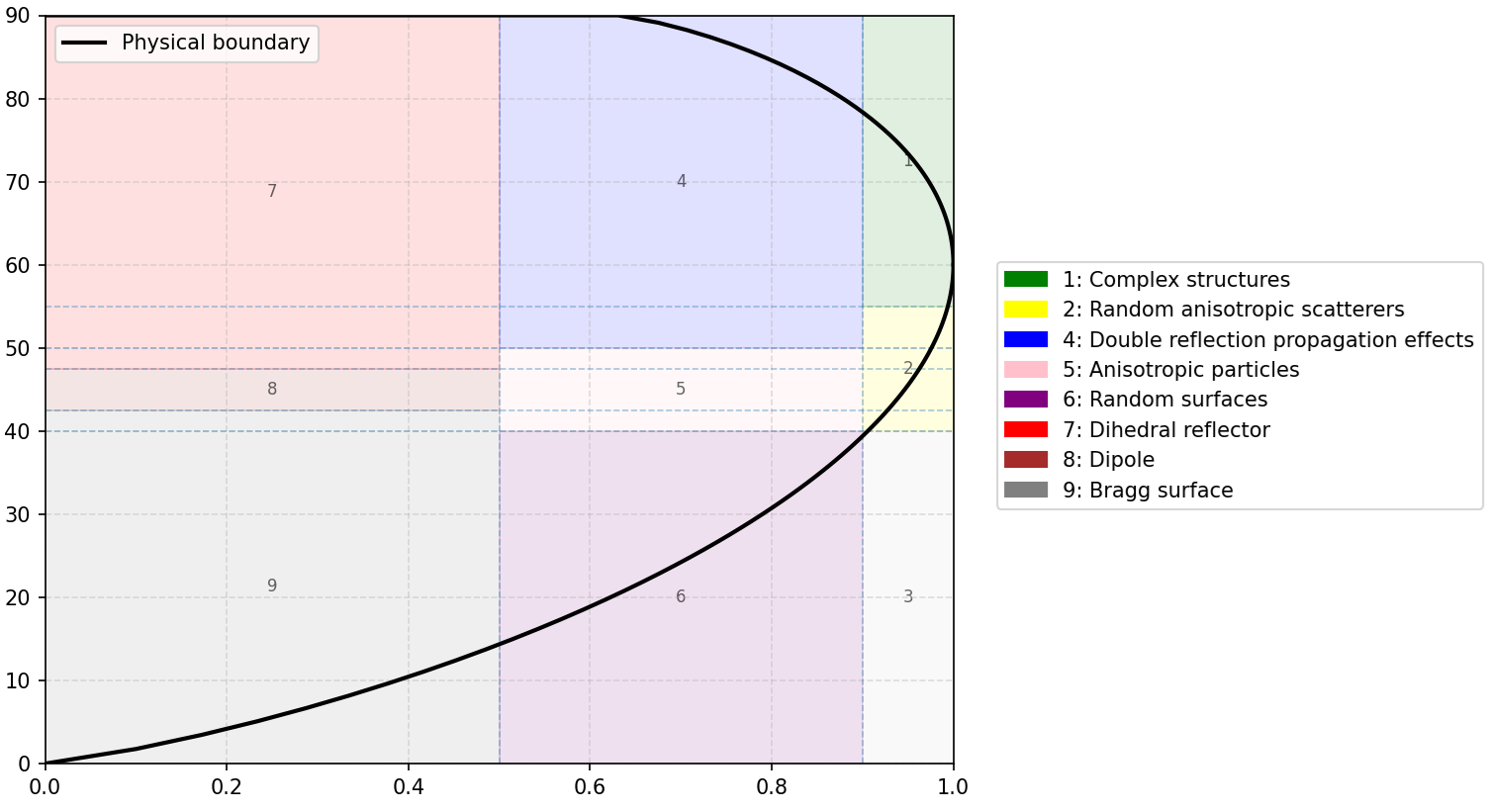}
    \caption{$H-\alpha$ plane separated into areas 1 to 9, each corresponding to a specific scattering mechanism, with the entropy at the x-axis and the scattering angle at the y-axis. The black line represents the boundary of physically possible $H-\alpha$ couples.}
    \label{fig:Halpha}
\end{figure}

The coefficients $k_s$, $k_d$, and $k_h$ represent the amplitude of each canonical scattering mechanism contributing to the originally measured scattering matrix $\mathbf{S}$. Similarly to the Pauli decomposition, we define the Krogager vector $\displaystyle \mathbf{h} = \left( k_{d}, k_{h}, k_{s} \right)^T$, which can also be assimilated to an RGB image.
\\
\\
\textbf{Cameron decomposition.} 
Unlike the Pauli and Krogager decompositions, the Cameron decomposition is not based on the principle of canonical scattering matrices but on two properties of radar targets: reciprocity and symmetry \cite{cameron1990feature, cameron2002simulated, cameron2006conservative}. The principle of reciprocity implies that $S_{hv}=S_{vh}$ in the linear polarization basis, vertical and horizontal. 
The reciprocity principle then divides the space of scattering matrices into two subspaces, one containing the Sinclair matrices of scatterers satisfying the reciprocity principle, the other containing the Sinclair matrices of non-reciprocal scatterers. The Cameron decomposition provides a coherent factorization of the polarimetric target in the Pauli domain as a sum of rank-1 canonical scatterers. Instead of expanding on a fixed analytical basis, it identifies a minimal set of elementary scattering vectors whose coherent superposition reproduces the measured response:
\begin{equation}
    \mathbf{k}_{\mathbf{S}} = \mathrm{vec}(\mathbf{S}) = \sum_{n=1}^{N} a_n\,\mathbf{v}_n ,
\end{equation}
where $\mathbf{v}_n$ are canonical single-mechanism vectors, $a_n$ are complex amplitudes, and $N$ is the number of physical scattering mechanisms actually needed to reconstruct the target.
Cameron does not impose a conventional set of mechanisms; instead, it reconstructs the target as a minimal sum of rank-1 scatterers, which may correspond to planes, dihedrals, helices, etc., but are selected based on the target itself rather than from a predefined basis. The key contribution is \emph{parsimony with physical closure}: the target is reconstructed using as few elementary coherent mechanisms as possible while preserving phase through the complex coefficients. This yields a compact and physically interpretable description of the scattering, revealing how canonical components combine coherently to produce the observed echo. In our setting, Cameron's decomposition finally leads to expressing the scattering matrix $\mathbf{k}_{\mathbf{S}}$ in terms of symmetrical contributions (trihedral, dihedral, narrow diplane, dipole, cylinder, quarter wave device), non-symmetrical contributions (left helix, right helix), and non-reciprocal contributions.
\\
\\
\textbf{$H-\alpha$ decomposition.}
\label{sec:halpha}
Although coherent decompositions help describe artificial targets, they are not suited to analyze random scattering effects (forests, fields, vegetation, etc.). To fill this lack, the Pauli vector $\mathbf{k}_{\mathbf{S}}$ is usually modeled by the multivariate, centered, circular complex Gaussian distribution $\mathcal{CN}(\mathbf{0}, \mathbf{T})$, which is fully characterized by the covariance matrix $\mathbf{T}=E\left[\mathbf{k}_{\mathbf{S}} \,\mathbf{k}_{\mathbf{S}}^H\right]$.
This covariance matrix is estimated locally on a PolSAR image through the Sample Covariance Matrix (SCM) $\hat{\mathbf{T}}$ computed on a set of $N$ samples taken in a spatial boxcar. This estimation can be performed through the well-known conventional Sample Covariance Matrix (SCM), defined as  
\begin{equation}
    \hat{\mathbf{T}} = \frac{1}{N} \sum_{i=1}^N \mathbf{k}_i\, \mathbf{k}_i^H\, .
    \label{eq:coherence}
\end{equation}
From the coherency matrix \(\mathbf{T}\), it's possible to perform a scattering decomposition, typically a superposition of different scattering contributions within the resolution cell. The goal is to decompose this complex signature into a sum of elementary scattering contributions, thereby better understanding the nature of the targets present in the radar image and extracting useful information. 
The Hermitian symmetric matrix $\hat{\mathbf{T}}$ can be diagonalized into $\lambda_1 \, \mathbf{e}_1 \, \mathbf{e}_1^H + \lambda_2 \, \mathbf{e}_2 \, \mathbf{e}_2^H + \lambda_3 \, \mathbf{e}_3 \, \mathbf{e}_3^H$ where $\left\{\lambda_i\right\}_{i\in\{1,2,3\}}$ and $\left\{\mathbf{e}_i\right\}_{i\in\{1,2,3\}}$ are the eigenvalues and eigenvectors respectively.
The entropy $H$ indicates the randomness of the overall backscattering phenomenon: $H = \displaystyle\sum_{i=1}^3  p_i \, \log{p_i}$ with the pseudo-probabilities  $p_i = \lambda_i /\left(\sum_{j=1}^3 \lambda_j \right) $ where $\lambda_i$ are the eigenvalues of $\hat{\mathbf{T}}$. Entropy $H$ varies between $0$ and $1$. A low entropy indicates that the observed target is pure and the backscattering is deterministic. This is reflected by a single non-zero normalized eigenvalue close to $1$. When the entropy is high, it reflects the completely random nature of the observed target. This occurs when the pseudo-probabilities are identical.
The angle $\alpha$ is defined as $\alpha =  \displaystyle\sum_{i=1}^{3} p_i\, \alpha_i$ where $\alpha_i = \arccos(|e_{i1}|)$ and where $e_{i1}$ is the first component of the $i$-th eigenvector of the SCM. It varies between $0$ and $\displaystyle\frac{\pi}{2}$ and characterizes the type of the dominant scattering mechanism (surface diffusion, dihedral diffusion). The relationship between entropy, $\alpha$ angle, and scattering mechanisms is represented in Figure~\ref{fig:Halpha}.
A classification procedure can be defined based on the entropy and the alpha parameters. Indeed, by considering the two-dimensional $H-\alpha$ space, all random scattering mechanisms can be represented \cite{formont2010statistical}. Therefore, a pixel belonging to a region of the $H-\alpha$ plane allows a physical interpretation of the average scattering mechanism.
\subsection{Complex-valued Convolutional Neural Networks}
\label{sec:cvnn}
Extending real-valued neural networks to the complex domain involves redefining some key components. In this article, we focus solely on Convolutional Neural Networks (CNNs), leaving the discussion open for other architectures, such as Vision Transformers (ViTs) \cite{Eilers2023}. The \cor Convolutional \fin AutoEncoder architecture introduced at the end of this section is made of a sequence of layers operating on the input representation. These layers include:
\begin{itemize}
\item Complex-valued convolution and linear layers, to extract new features by combining the lower-level features,
\item  Complex-valued activations, building blocks for a non-linear predictor,
\item Complex-valued pooling layers, which reduce the spatial dimensions of the representation,
\item Complex-valued up-sampling layers, which interpolate the input features to enlarge the spatial dimensions of the representation.
\end{itemize}
Complex-valued normalization layers have also been introduced and shown, in the real case, to speed up the optimization and possibly introduce some regularization. Complex-valued gradient descent and initialization are discussed at the end of this section.
\\
\\
\textbf{Complex-valued cross‐correlation layers.}
Cross‐correlation layers are naturally extended to the complex domain, given the definition of the complex-valued convolutional layer: for a given complex input matrix $\mathbf{Z}$, the complex-valued cross‐correlation operation is defined as:
\begin{equation}
    \mathrm{Conv}(\mathbf{Z})(i,j) = \displaystyle\sum_{p} \sum_{q} \mathbf{Z}(i+p, j+q) \, F(p, q) \, ,
\end{equation}
where $F$ is a kernel, \((i, j)\) the coordinates inside of the matrix, and $p$ and $q$ iterate over the filter dimensions. Note that we keep the $\mathrm{Conv}$ notation that is widely used in modern CNNs, even if the operation is in fact not a convolution.
\\
\\
\textbf{Complex-valued linear layers.}
Based on the definition of the real-valued dense layer and the definition of the complex-valued neuron, we can define a complex-valued dense layer in the form of a tensor, such as:
\begin{equation}
    \mathrm{Linear}(\mathbf{z})=\mathbf{W}\,\mathbf{z}+ \boldsymbol{\beta}  \, ,
\end{equation}
where $\mathbf{W}\in \mathbb{C}^{m \times n}$ is the weight matrix and $\boldsymbol{\beta} \in \mathbb{C}^m$ is a bias vector. These linear layers can be used in the bottleneck of the AutoEncoder. 
\\
\\
\textbf{Complex-valued activations.}
Several activation functions have been investigated in the literature. Kuroe et al. \cite{kuroe2003activation} proposed to categorize activation functions into the following two classes of complex functions $f(\cdot)$:
\begin{equation}
\begin{array}{l}
    \text{Type A}: f(z) = f_{\Re}\left(\Re(z)\right) + j\, f_{\Im}\left(\Im(z)\right)\, , \\ \\
    \text{Type B}: f(z) = \psi(r)\, e^{j\phi(\theta)}\, ,
    \end{array}
\end{equation}
where $f_{\Re}: \mathbb{R}\mapsto\mathbb{R}$ and $f_{\Im}: \mathbb{R}\mapsto\mathbb{R}$ are nonlinear real-valued functions. $\psi: \mathbb{R}^{+}\mapsto\mathbb{R}^{+}$ and $\phi: ]-\pi; \pi]\mapsto\mathbb{R}$ are also nonlinear real functions. Type-A and Type-B functions differ in how they represent the complex numbers on which they operate. Type-A activation functions apply a real-valued function on real and imaginary parts. These functions can be the same, although they do not have to. Type-B activation functions operate independently on the magnitude and phase of the complex numbers. 
We consider some of the most well-known activation functions for complex-valued neural networks. The $\mathrm{modReLU}(\cdot)$ function defined is as \cite{arjovsky2016unitary}:
\begin{eqnarray}
    \mathrm{modReLU}(\mathbf{z}) & = &\mathrm{ReLU}\left(\left|\mathbf{z}\right|+b\right)\, e^{j\theta} \nonumber \\
    &= &
    \begin{cases}
    \left(\left|\mathbf{z}\right|+b\right)\displaystyle\frac{\mathbf{z}}{\left|\mathbf{z}\right|} & \text{if}\ \left|\mathbf{z}\right|+b>0\, , \\
      0 & \text{otherwise}\, .
    \end{cases}
\end{eqnarray}
where the learnable parameter $\boldsymbol{\theta}$ is the phase vector of each component of the vector $\mathbf{z}/|\mathbf{z}|$ and  $b$ is a bias characterizing a "dead zone" around the origin where the neuron is inactive.  The $\mathrm{zReLU}(\cdot)$  function is defined as \cite{guberman2016complex, barrachina2023theory}:
\begin{eqnarray}
    \begin{split}\mathrm{zReLU}(\mathbf{z}) = \begin{cases} \mathbf{z} \quad \text{if} \quad \boldsymbol{\theta} \in [0, \pi/2] \\ 0 \quad \text{else} \end{cases}\end{split}
\end{eqnarray}
where the learnable parameter $\boldsymbol{\theta}$ is the phase vector of each component of the vector $\mathbf{z}/|\mathbf{z}|$. The $\mathrm{CReLU}(\cdot)$, which is the function used in this paper, is defined as \cite{gao2018enhanced}:
\begin{eqnarray}
    \mathrm{CReLU}(\mathbf{z}) = \mathrm{ReLU}(\mathbf{x}) + j \,\mathrm{ReLU}(\mathbf{y})
\end{eqnarray}
Finally, the $\mathrm{Cardiod}(\cdot)$ function is defined as \cite{virtue2017better, barrachina2023theory}:
\begin{eqnarray}
    \mathrm{Cardiod}(\mathbf{z}) = \frac{1}{2} (1 + \text{cos}(\boldsymbol{\theta})) \odot \mathbf{z}
\end{eqnarray}
where the learnable parameter $\boldsymbol{\theta}$ is the phase vector of each component of the vector $\mathbf{z}/|\mathbf{z}|$. 
\\
\\
\textbf{Complex-valued down-sampling \& pooling layer.} Commonly used pooling layers are fundamentally a filter operation (max, mean, etc.) with a stride of $1$ followed by a sub-sampling (using a stride strictly larger than $1$). The filtering operation is applied on a window of a specific size. A down-sampling-by-$p$ factor operator $\mathbf{D}_p:  \mathbb{C}^{N} \mapsto \mathbb{C}^{ \left\lfloor \frac{N}{p} \right\rfloor}$ is defined as:
\begin{equation}
    \mathbf{D}_p(\mathbf{z})[n]=\mathbf{z}[p \, n] \, \, \, , \ \forall n \in \left[ 1, \left\lfloor \frac{N}{p} \right\rfloor\right].
    \label{eq:downsampling}
\end{equation}
Note that some pooling layers in the complex domain differ from their real-valued counterparts. The max pooling layer cannot be trivially extended to the complex domain as $\mathbb{C}$ is only partially ordered. As such, the returned element is a complex-valued scalar such that the selection rule is ordered, the complex-valued max pooling layer $\mathrm{MaxPool}: \mathbb{C}^{N} \mapsto \mathbb{C}$ can be defined as:
\begin{equation}
    \mathrm{MaxPool}(\mathbf{z})=\mathbf{z}[k] \mbox{ where } k = \arg \max_{i\in [1,n]}\left|\mathbf{z} [i]\right| \, ,
\end{equation}
Similarly, the complex-valued average pooling layer $\mathrm{AvgPool}:\mathbb{C}^{N} \mapsto \mathbb{C}$ is defined as:
\begin{equation}
    \mathrm{AvgPool}(\mathbf{z}) = \mathrm{AvgPool}(\mathbf{\Re(z)}) + j \, \mathrm{AvgPool}(\mathbf{\Im(z)})= \frac{1}{N} \sum_{i=1}^{N} \left(\Re(\mathbf{z}[i]) + j \, \Im(\mathbf{z}[i])\right)\, ,
\end{equation}
where $\mathbf{z}=(\mathbf{z}[1], \ldots, \mathbf{z}[N])^T \in \mathbb{C}^{N}$, and $N$ is the number of elements in the pooling window. By applying the $\mathrm{AvgPool}$ layer separately on the real and imaginary parts, we ensure that the structure of the complex signal is preserved throughout the down-sampling process. For the model presented in the article, we select a down-sampling-by-$2$ factor operator which is included in a $\mathrm{Conv}$ layer, without any pooling filter operation.
\\
\\
\textbf{Complex-valued up-sampling.}
While the down-sampling/pooling layers compress the spatial dimensions of their input representations, the expansion of the spatial dimensions is obtained with up-sampling layers. Several propositions have been made in the literature for this expansion. The early approach used in \cite{ronneberger2015u} employed transposed convolution (also known as fractionally strided convolution), but this proved to introduce checkerboard artifacts~\cite{Odena2016} if the hyperparameters were not chosen appropriately. An alternative approach, used in this paper, is to apply a non-trainable  $\mathrm{Nearest}$  interpolation layer, followed by trainable convolutions.
\\
\\
\textbf{Complex-valued normalization layers.}
In real-valued neural networks, introducing normalization layers has been demonstrated to accelerate optimization~\cite{Ioffe2015}. Various normalization layers have been introduced, beginning with BatchNorm~\cite{Ioffe2015}, followed by LayerNorm~\cite{Ba2016}, and RMSNorm~\cite{Zhang2019}. These layers differ primarily in how they compute normalization statistics: BatchNorm computes them over a batch of samples, whereas LayerNorm and RMSNorm compute them over the features of a single sample. Additionally, RMSNorm distinguishes itself from LayerNorm by omitting the centering step. 
\\
\\
Standardizing an array $\mathbf{z}$ of complex numbers is not as straightforward as scaling them to achieve a zero mean and unit variance. This approach does not ensure equal variance in both the real and imaginary components (non-circularity property). Trabelsi et al. \cite{trabelsi2017deep} proposed an extension of the BatchNorm layer to the complex domain, taking into account the two-dimensional nature of the input. They proposed to characterize each complex component $\mathbf{z}[i]$ of the vector $\mathbf{z}$ as a bi-dimensional real vector $\mathbf{x}_i = \left(\Re(\mathbf{z}[i]),\Im(\mathbf{z}[i])\right)^T$. This vector is then centered and whitened and characterizes the vector $\tilde{\mathbf{x}}_i = \left(\Re\left(\tilde{\mathbf{z}}[i]\right), \Im\left(\tilde{\mathbf{z}}[i]\right)\right)^T$ according:
\begin{equation}
\tilde{\mathbf{x}}_i = 
    \begin{pmatrix}
\text{Cov}\left(\Re\left(\mathbf{z}[i]\right), \Re\left(\mathbf{z}[i]\right)\right) & \text{Cov}\left(\Re\left(\mathbf{z}[i]\right), \Im\left(\mathbf{z}[i]\right)\right) \\
\text{Cov}\left(\Im\left(\mathbf{z}[i]\right), \Re\left(\mathbf{z}[i]\right)\right) & \text{Cov}\left(\Im\left(\mathbf{z}[i]\right), \Im\left(\mathbf{z}[i]\right)\right)
    \end{pmatrix}^{-\frac{1}{2}} \left( \mathbf{x}_i - E\left[\mathbf{x}_i\right] \right)\, .
\end{equation}
This operation ensures that $\mathbf{z}[i]$ has a standard complex Normal distribution with mean $E\left[\tilde{\mathbf{z}}[i]\right]=0$, identity variance $E\left[\tilde{\mathbf{z}}[i]\, \tilde{\mathbf{z}}[i]^\ast\right]=1$, and pseudo-variance equal to $E\left[\tilde{\mathbf{z}}[i]\, \tilde{\mathbf{z}}[i]^T\right]=0$. Thus, the real and imaginary parts of $\tilde{\mathbf{z}}[i]$ are uncorrelated. Analogously to the real-valued batch normalization algorithm, we define the $\mathrm{BatchNorm}(\cdot)$ operator as:
\begin{equation}
    \mathrm{BatchNorm}(\tilde{\mathbf{x}_i}) =
    \begin{pmatrix}
    Re(\tilde{\mathbf{z}}[i])\\
    \Im(\tilde{\mathbf{z}}[i])
    \end{pmatrix} = 
    \boldsymbol{\beta} + \boldsymbol{\Gamma} \, \tilde{\mathbf{x}}_i\, ,
\end{equation}
where $\boldsymbol{\beta}$ and $\boldsymbol{\Gamma}$ are respectively a learnable real-valued vector and a learnable $2\times 2$ real-valued matrix of parameters. The complex batch normalization $\tilde{\mathbf{z}}$ of the complex vector $\mathbf{z}$ is then defined by the vector characterized by components $\left\{\tilde{\mathbf{z}}[i]\right\}_i$ equal to the complex recombination of the bi-dimensional vector $\mathrm{BatchNorm}(\tilde{\mathbf{x}_i})$ components.
\\
\\
\textbf{Complex-valued \cor Convolutional \fin AutoEncoders.}
As explained by Rumelhart  \cite{rumelhart_learning_1986}, an AutoEncoder (AE) is a neural network that encodes the input in a latent space and reconstructs its input signal via a decoder. Consequently, when dealing with spatial data types like images, encoding and decoding are performed using convolutional layers; such models are referred to as Convolutional \cor AutoEncoders \fin (CoAEs) \cite{honkela_stacked_2011}.
In this work, we extend the \cor CoAE \fin architecture to complex-valued data, following the approach of Asiyabi et al. \cite{asiyabi2022complex}. The AE structure usually includes three parts (with the latent space being implicitly defined in some cases), as illustrated in Figures \ref{fig:coae} and \ref{fig:colae}:
\begin{figure}
    \centering
    \includegraphics[width=0.75\linewidth]{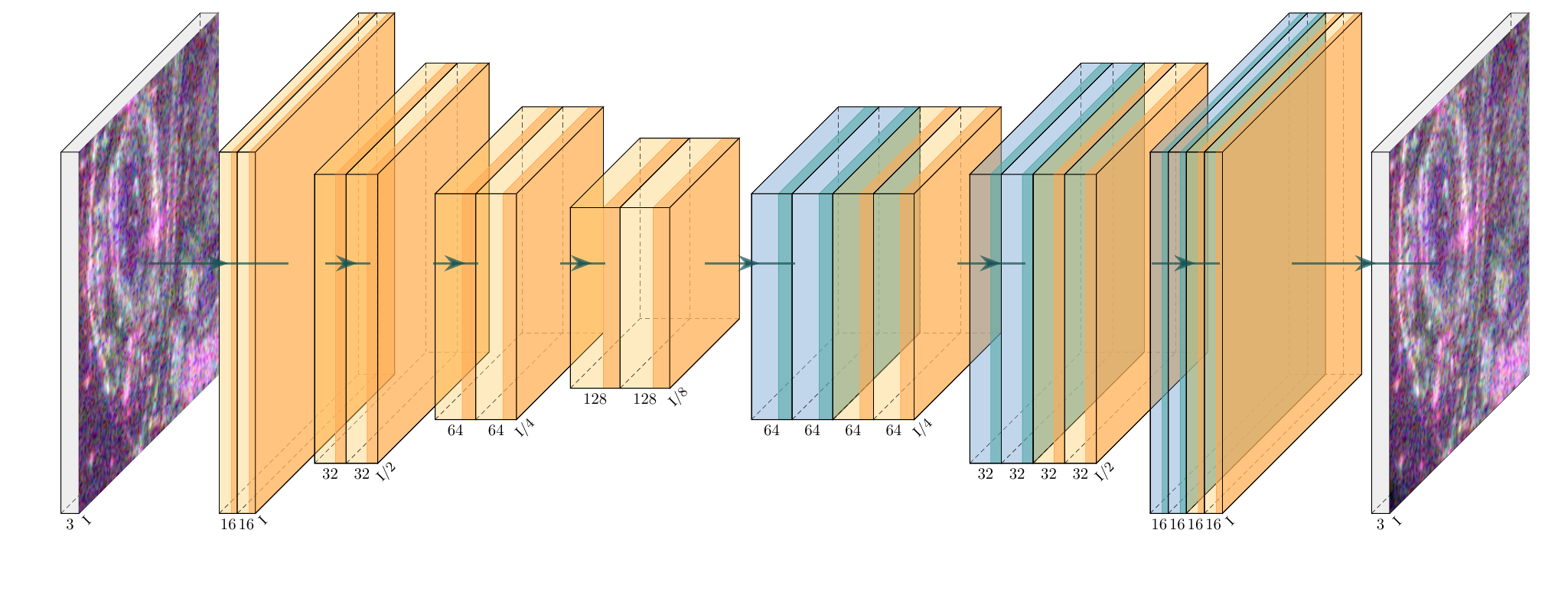}
    \caption{Architecture of a Convolutional AutoEncoder with the complex-valued residual blocks (yellow), and the complex-valued up-sampling layers (blue).}
    \label{fig:coae}
\end{figure}
\begin{figure}
    \centering
    \includegraphics[width=0.75\linewidth]{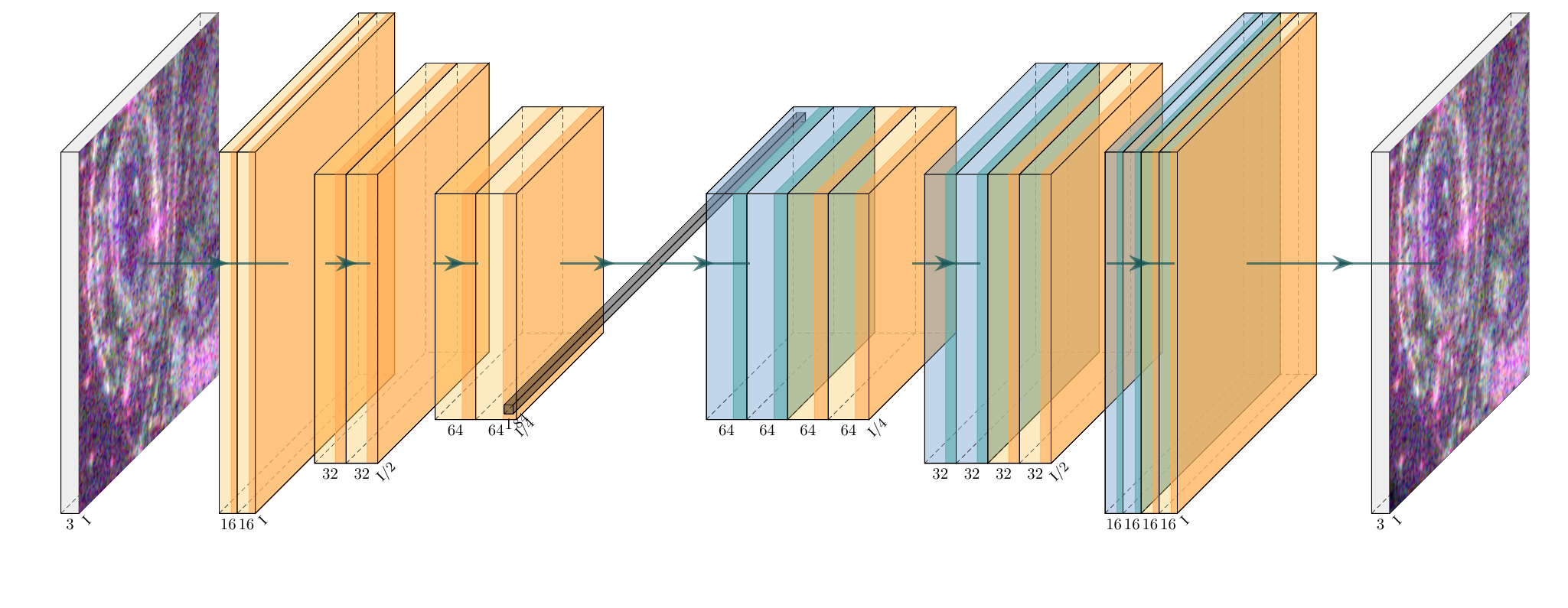}
    \caption{Architecture of a Convolutional AutoEncoder with an explicit Bottleneck. The same color code as in Figure~\ref{fig:coae} applies with the additional dense layers between the encoder and decoder.}
    \label{fig:colae}
\end{figure}
\begin{itemize}
    \item $\mathrm{Encoder}$: compression of the input data $\mathbf{X}$ by progressively reducing its dimensionality. In the case of a CoAE, it is a succession of convolutional layers, activation functions, and spatial down-sampling. More precisely, we define our encoder as an input $\mathrm{Conv}$ layer, followed by a sequence of $\mathrm{ResBlock}$. The spatial dimension is reduced in each $\mathrm{ResBlock}$ layer throughout the $\texttt{Encoder}$. As such, we can express this operation compactly as~:
    \begin{equation}
        \mathrm{Encoder}(\mathbf{X}) = \mathrm{ResBlock}_{N}(\cdots(\mathrm{ResBlock}_{0}(\mathrm{Conv}(\mathbf{X}))))
    \end{equation}
    We can further define a $\mathrm{ResBlock}$ as a succession of two $\mathrm{Conv}$ layers, adding the output of the second $\mathrm{Conv}$ to the input of the first one. These residual connections, which are at the core of the ResNet architecture~\cite{he2016deep}, have been shown to ease the propagation of the gradient, fighting against the vanishing gradient problem. Additionally, each of them is followed by a $\mathrm{modReLU}$ and a $\mathrm{BatchNorm}$. Thus, we can express the $\mathrm{ResBlock}$ as:
    \begin{align}
        Y &= \mathrm{BatchNorm}(\mathrm{modReLU}(\mathrm{Conv}(X)))) \\
        Y' &= X + \mathrm{BatchNorm}(\mathrm{modReLU}(\mathrm{Conv}(Y))))
    \end{align}
    \item $\mathrm{Bottleneck}$: it may be explicitly expressed as a flattened dense layer of dimension $p$. In this case, we apply a $\mathrm{Flatten}$ operator to obtain a spatially flattened vector (with a channel dimension). We apply a $\mathrm{Linear}$ layer to reduce the flattened vector to a given latent dimension, then a $\mathrm{BatchNorm}$ layer, and a $\mathrm{modReLU}$ activation function. The process is performed in the opposite way to recover the latent vector to its initial size, before applying an $\mathrm{Unflatten}$ operator to reintroduce the spatial dimensions that were lost at the start of the bottleneck. As such, we can express the bottleneck in its most summarized form as: \begin{align}
        Y &= \mathrm{BatchNorm}(\mathrm{modReLU}(\mathrm{Linear}_{0}(\mathrm{Flatten}(\mathbf{X})))) \\
        \mathrm{Bottleneck}(\mathbf{X}) &= \mathrm{Unflatten}(\mathrm{BatchNorm}(\mathrm{modReLU}(\mathrm{Linear}_{1}(Y))))
    \end{align}
    \item $\mathrm{Decoder}$: it expands the latent vector back to the original spatial dimensions, resulting in a reconstructed data $\hat{\mathbf{X}}$. In the case of a CoAE, the decoder is a succession of spatial up-sampling, convolutional layers, and activation functions. The decoder essentially mirrors the encoder, except that it replaces the down-sampling layer with the up-sampling layer  $\mathrm{Nearest}$.  
    As such, we can express the $\mathrm{Decoder}$ in its most summarized form as: \begin{equation}
        \mathrm{Decoder}(\mathbf{X}) = \mathrm{ResBlock}_{N}(\cdots(\mathrm{ResBlock}_{0}(\mathrm{Nearest}(\mathrm{Bottleneck}(\mathbf{X})))))
    \end{equation}
\end{itemize}
\textbf{Complex differentiability.}
The partially ordered nature of $\mathbb{C}$ is incompatible with minimization and optimization. Thus, loss functions $L$, estimating the correctness of the model prediction, must return a real-valued scalar ($L:\mathbb{C}^{n} \mapsto \mathbb{R}$). One common way to define such functions, like the Cross Entropy Loss, is to consider only the magnitude of $\mathbf{z}$ or compute the loss functions on the real and imaginary parts separately. As we can deduce from this discussion, the projection from $\mathbb{C}$ to $\mathbb{R}$ is still an open question, with no consensus.
\\
Additionally, on the topic of complex differentiability, since the loss $L$ is bounded, it is not holomorphic according to Liouville's theorem. Therefore, the Wirtinger derivative is used to compute the gradient of the loss $L$ with respect to the real and imaginary parts of the neural network weights in the complex backpropagation algorithm \cite{sarroff2015learning}.
\\
\\
\textbf{Complex-valued initialization.} 
Gradient descent, using Wirtinger derivatives, is an iterative process. Part of the success of deep learning is due to the careful choice of initialization functions. The adaptation of the two most well-known real-valued normalization layers (Xavier \cite{glorot_understanding_2010}, and He \cite{he_delving_2015}) relies on the fact that the variance of the real and imaginary parts is half the variance of the entire complex distribution \cite{trabelsi2017deep}. Let us first define the complex-valued Xavier initialization:
\begin{align}
    \text{Re}(\omega)&=\text{Im}(\omega) \sim \displaystyle \mathcal{N}\left(0, \sqrt{\frac{1}{{n_{\text{in}} + n_{\text{out}}}}}\right) \, ,\\
    \text{Re}(\omega)&=\text{Im}(\omega) \sim \mathcal{U}\left[-\displaystyle\sqrt{\frac{3}{{n_{\text{in}} + n_{\text{out}}}}}, \sqrt{\frac{3}{{n_{\text{in}}}}} \right]\, .
\end{align}
where $n_{in}$ and $n_{out}$ are respectively the fanin and fanout of the units. Then, let us define the complex-valued He initialization:
\begin{align}
    \text{Re}(\omega)&=\text{Im}(\omega) \sim \displaystyle \mathcal{N}\left(0, \sqrt{\frac{1}{{n_{\text{in}}}}}\right) \, ,\\
    \text{Re}(\omega)&=\text{Im}(\omega) \sim \mathcal{U}\left[-\displaystyle\sqrt{\frac{3}{{n_{\text{in}}}}}, \sqrt{\frac{3}{{n_{\text{in}}}}} \right]\, .
\end{align}
\section{Experiments}
\label{sec:experiments}
In this section, we assess the performance of the proposed complex-valued Convolutional AutoEncoder architecture for PolSAR image reconstruction, with particular emphasis on the preservation of polarimetric properties. Section \ref{sec:ablation} presents a comprehensive ablation study analyzing the contribution of key architectural components to the overall performance. Section \ref{sec:res} further compares our approach against dual real-valued neural networks (dual-RVNNs) to evaluate the benefits of using complex-valued representations in contrast to their real-valued counterparts. \cor It is worth noting that we excluded amplitude-only AutoEncoders from this study. Since coherent decompositions (such as Pauli and Cameron) and the $H-\alpha$ analysis fundamentally rely on the relative phase between polarization channels, amplitude-only models are theoretically incapable of preserving these physical properties. Consequently, the dual-RVNN serves as the most rigorous baseline, as it retains access to the full complex signal information, albeit processed in the real domain. \fin
\\
We extend the evaluation procedure introduced by Gabot et al. \cite{gabotpreserving} to assess the preservation of polarimetric properties. First, we compute reconstruction metrics on the test set, including the Mean Squared Error (MSE), Peak Signal-to-Noise Ratio (PSNR), and Structural Similarity Index Measure (SSIM). While MSE and PSNR are computed on the complex-valued data, SSIM is evaluated between the amplitude of the input and reconstructed images. Subsequently, we perform polarimetric decompositions on the full original and reconstructed images to analyze the preservation of polarimetric characteristics. Our analysis focuses specifically on the Cameron and $H-\alpha$ decompositions, as they provide classification-based evaluation metrics. From these decompositions, we derive the Accuracy (Acc) and F1-Score (F1) values. We emphasize that all reported metrics are computed in an unsupervised manner on a model trained for reconstruction and are therefore not utilized during the training process.
\\
\\
We provide a complete guide to reproduce our experiments at the following repository \footnote{\url{https://github.com/QuentinGABOT/Exploring-polarimetric-properties-preservation-with-CVNNs}}.
\subsection{Datasets}
\textbf{San Francisco Polarimetric SAR ALOS-2.}
The San Francisco Polarimetric SAR ALOS-2 dataset\footnote{https://ietr-lab.univ-rennes1.fr/polsarpro-bio/san-francisco/} is an open-access PolSAR dataset. Specifically, it consists of an L-band polarimetric SAR image acquired by the ALOS-2 satellite, featuring an approximate ground range resolution of $10\,$m. Owing to the penetration capability of L-band waves through forest canopies, vegetation, snow, and soil, ALOS-2 provides valuable information about surface and subsurface structures of the Earth \cite{yamaguchi2016alos}. The San Francisco dataset is fully polarimetric, meaning that its four channels correspond to the four elements of the Sinclair matrix. \\
The dataset is built by cropping the $22,608 \times 8,080$ image into $64\times 64$ non-overlapping tiles, resulting in a dataset of $44,478$ four-channel tiles. Tiles are randomly assigned to the training, validation, and test folds with $80\%$ for training, $10\%$ for validation, and $10\%$ for test.
Note that the image in the experiments is a crop of the original image to reduce computation cost, resulting in a $4200 \times 2000$ image.
\\
\\
\textbf{Brétigny.}
The Brétigny dataset is a three-channel PolSAR dataset acquired by the RAMSES sensor \cite{Ramses}. It features a spatial resolution of $1.32\,$m in range and $1.38\,$m in azimuth, with an incidence angle of $30\deg$ and operating in the X-band frequency range \cite{formont2010statistical}. The original $1536 \times 3392$ image is divided into non-overlapping tiles of size $64 \times 64$, yielding a total of 1224 three-channel samples. These tiles are randomly partitioned into training, validation, and test sets, using an 80\%/10\%/10\% split, respectively.
\begin{table}[htbp]
    \centering
    \cor
    \begin{tabular}{ll}
        \toprule
        \multicolumn{2}{c}{\textbf{Model Architecture}} \\
        \midrule
        \textbf{Parameter} & \textbf{Value/Method} \\
        \midrule
        Input Image Size & $64 \times 64 \times 4$ \\
        Encoder/Decoder Depth & $2$ \\
        Initial Channel Width & $64$ \\
        Convolutional Kernel Size & $3\times3$ \\
        Downsampling Method & Strided Convolution \\
        Upsampling Method & Nearest Neighbor + Conv \\
        Activation Function & CReLU ($\mathbb{C}$) / ReLU ($\mathbb{R}$) \\
        Weight Initialization & Complex He ($\mathbb{C}$) / He ($\mathbb{R}$) \\
        Normalization Layer & Complex Batch Normalization ($\mathbb{C}$) / Batch Normalization ($\mathbb{R}$)\\
        \midrule
        \multicolumn{2}{c}{\textbf{Training Configuration}} \\
        \midrule
        \textbf{Parameter} & \textbf{Value} \\
        \midrule
        Optimizer & AdamW \\
        Learning Rate & $5e-4$ \\
        Weight Decay & $1e-3$ \\
        Batch Size & $32$ \\
        Total Epochs & $250$ \\
        Loss Function & Mean Squared Error (MSE) \\
        Implementation Framework & PyTorch / torchcvnn \\
        Hardware & NVIDIA A100 \\
        \bottomrule
    \end{tabular}
    \caption{\cor Summary of the complex-valued Convolutional AutoEncoder architecture and training hyperparameters. Note that for the CVNN, all weights and operations are performed in the complex domain $\mathbb{C}$, whereas the RVNN baseline uses the equivalent real-valued setup.\fin}
    \fin
    \label{tab:hyperparameters}
\end{table}
\subsection{Ablation study}
\label{sec:ablation}
We evaluate the influence of three key architectural design choices: activation nonlinearity, network depth, and down/upsampling operators. Models were trained using the AdamW optimizer \cite{loshchilov2017decoupled} to minimize the Mean Squared Error (MSE) reconstruction loss. We employed a complex-valued CoAE architecture, as detailed in Fig.~\ref{fig:coae}. \cor The complete architecture specifications and specific training hyperparameters (learning rate, batch size, etc.) are detailed in Table~\ref{tab:hyperparameters}. \fin Except for the factor under ablation, all other training settings and hyperparameters were kept constant.
\\
\\
\textbf{Activation functions (Table~\ref{tab:activation}).}
\begin{table}[htbp]
\centering
\resizebox{\linewidth}{!}{
\begin{tabular}{|c|c|c|c|c|c|c|c|c|c|c|c|}
\hline
Activation function & Dataset & MSE $(\downarrow)$ & PSNR $(\uparrow)$ & SSIM $(\uparrow)$ & $H$ –- $\alpha$ OA (\%) $(\uparrow)$ & $H$ –- $\alpha$ F1 (\%) $(\uparrow)$ & Cam. OA (\%) $(\uparrow)$ & Cam. F1 (\%) $(\uparrow)$ \\
\hline
$\boldsymbol{\mathrm{CReLU}}$ & ALOS-2 & $\mathbf{1.1\times10^{-3}}$ & $\mathbf{29.77}$ & $\mathbf{0.99}$ & $93.81$ & $93.80$ & $\mathbf{88.03}$ & $\mathbf{88.05}$\\
\hline
$\mathrm{Cardioid}$ & ALOS-2 & $1.2\times10^{-3}$ & $29.05$ & $\mathbf{0.99}$ & $\mathbf{93.87}$ & $\mathbf{93.82}$ & $87.22$ & $87.25$ \\
\hline
$\mathrm{modReLU}$ & ALOS-2 & $8.2\times10^{-3}$ & $20.86$ & $0.92$ & $86.43$ & $86.35$ & $71.77$ & $71.98$ \\
\hline
$\mathrm{zReLU}$ & ALOS-2 & $1.7\times10^{-2}$ & $17.72$ & $0.85$ & $81.45$ & $81.46$ & $69.22$ & $69.49$ \\
\hline
\hline\hline
$\boldsymbol{\mathrm{CReLU}}$ & Bretigny & $\mathbf{1.3\times10^{-3}}$ & $\mathbf{28.71}$ & $\mathbf{0.98}$ & $\mathbf{95.74}$ & $\mathbf{95.72}$ & $\mathbf{88.98}$ & $\mathbf{88.99}$ \\
\hline
$\mathrm{Cardioid}$ & Bretigny & $1.7\times10^{-3}$ & $27.50$ & $0.97$ & $95.51$ & $95.49$ & $87.11$ & $87.13$ \\
\hline
$\mathrm{modReLU}$ & Bretigny & $6.1\times10^{-3}$ & $21.90$ & $0.91$ & $91.15$ & $91.04$ & $77.45$ & $77.57$ \\
\hline
$\mathrm{zReLU}$ & Bretigny & $1.8\times10^{-2}$ & $17.53$ & $0.80$ & $83.89$ & $83.27$ & $67.67$ & $68.05$ \\
\hline
\end{tabular}
}
\caption{Reconstruction metrics, $H-\alpha$ classification metrics, and Cameron classification metrics are reported for the San Francisco Polarimetric SAR ALOS-2 and Brétigny datasets. Results for the ALOS-2 dataset are separated from those for Brétigny by an additional row. Across both datasets, the $\mathrm{CReLU}$ activation consistently outperforms other activation functions on nearly all metrics.}
\label{tab:activation}
\end{table}
The $\mathrm{CReLU}$ activation consistently outperforms alternatives across both reconstruction and downstream polarimetric metrics. On the ALOS-2 dataset, it achieves an approximately $7.5\times$ lower MSE compared to $\mathrm{modReLU}$ ($1.1\times10^{-3}$ vs.\ $8.2\times10^{-3}$), improves PSNR by $+8.91\,$dB ($29.77\,$dB vs. $20.86\,$dB), and increases Cameron OA by +16.3 percentage points (88.03\% vs.\ 71.77\%). On the Brétigny dataset, similar trends are observed, with an MSE reduction of approximately $4.7\times$ and a PSNR gain of $+6.81\,$dB. The $\mathrm{Cardioid}$ activation performs comparably to $\mathrm{CReLU}$ on reconstruction (MSE $\approx$ 9\% higher; PSNR $-0.72\,$dB) and occasionally matches it on $H-\alpha$ overall accuracy and F1-score (differences $\leq$ 0.1 pp). In contrast, $\mathrm{zReLU}$ exhibits severe degradation (MSE roughly $15\times$ higher than $\mathrm{CReLU}$ on ALOS-2 and $14\times$ on Brétigny), along with notable declines in SSIM and classification performance.
\\
\\
\textbf{Depth (Table~\ref{tab:depth}).}
\begin{table}[htbp]
\centering
\resizebox{\linewidth}{!}{
\begin{tabular}{|c|c|c|c|c|c|c|c|c|c|c|c|}
\hline
Model's depth & Dataset & MSE $(\downarrow)$ & PSNR $(\uparrow)$ & SSIM $(\uparrow)$ & $H$ –- $\alpha$ OA (\%) $(\uparrow)$ & $H$ –- $\alpha$ F1 (\%) $(\uparrow)$ & Cam. OA (\%) $(\uparrow)$ & Cam. F1 (\%) $(\uparrow)$ \\
\hline
$\boldsymbol{2}$ & ALOS-2 & $\mathbf{1.1\times10^{-3}}$ & $\mathbf{29.77}$ & $\mathbf{0.99}$ & $\mathbf{93.81}$ & $\mathbf{93.80}$ & $\mathbf{88.03}$ & $\mathbf{88.05}$\\
\hline
$3$ & ALOS-2 & $6.5\times10^{-3}$ & $21.86$ & $0.94$ & $88.34$ & $88.39$ & $75.21$ & $75.20$ \\
\hline
$4$ & ALOS-2 & $3.2\times10^{-1}$ & $4.89$ & $0.18$ & $45.16$ & $46.42$ & $25.06$ & $25.03$ \\
\hline
\hline\hline
$\boldsymbol{2}$ & Bretigny & $\mathbf{1.3\times10^{-3}}$ & $\mathbf{28.71}$ & $\mathbf{0.98}$ & $\mathbf{95.74}$ & $\mathbf{95.72}$ & $\mathbf{88.98}$ & $\mathbf{88.99}$ \\
\hline
$3$ & Bretigny & $1.8\times10^{-1}$ & $7.45$ & $0.33$ & $75.49$ & $75.32$ & $49.40$ & $49.76$ \\
\hline
$4$ & Bretigny & $3.1\times10^{-1}$ & $5.17$ & $0.10$ & $36.95$ & $34.39$ & $28.39$ & $26.79$ \\
\hline
\end{tabular}
}
\caption{Reconstruction metrics, $H-\alpha$ classification metrics, and Cameron classification metrics are reported for the San Francisco Polarimetric SAR ALOS-2 and Brétigny datasets. Results for the ALOS-2 dataset are separated from those for Brétigny by an additional row. Across both datasets, a model depth of two consistently outperforms the other depth variations.}
\label{tab:depth}
\end{table}
Increasing the network depth beyond two layers results in a significant degradation in performance. On the ALOS-2 dataset, a depth of three results in an approximately $5.9\times$ higher MSE and a PSNR decrease of $7.91\,$dB, while a depth of four causes catastrophic failure ($\sim 2.9\times 10^{2}$ increase in MSE; PSNR = $4.89\,$dB). On the Brétigny dataset, the effect is even more pronounced, with depth 3 yielding an MSE roughly $138\times$ higher and a PSNR drop of $21.26\,$dB; deeper models perform worse still. The $H-\alpha$ and Cameron metrics exhibit similar declines, reflecting a consistent collapse in both reconstruction and polarimetric quality.
\\
\\
\textbf{Down/upsampling (Table~\ref{tab:sampling}).}
\begin{table}[htbp]
\centering
\resizebox{\linewidth}{!}{ 
\begin{tabular}{|c|c|c|c|c|c|c|c|c|c|c|c|}
\hline
Downsampling layer & Upsampling layer & Dataset & MSE $(\downarrow)$ & PSNR $(\uparrow)$ & SSIM $(\uparrow)$ & $H$ –- $\alpha$ OA (\%) $(\uparrow)$ & $H$ –- $\alpha$ F1 (\%) $(\uparrow)$ & Cam. OA (\%) $(\uparrow)$ & Cam. F1 (\%) $(\uparrow)$ \\
\hline
\textbf{Strided Conv} & \textbf{Nearest} & ALOS-2 & $\mathbf{1.1\times10^{-3}}$ & $\mathbf{29.77}$ & $\mathbf{0.99}$ & $\mathbf{93.81}$ & $\mathbf{93.80}$ & $\mathbf{88.03}$ & $\mathbf{88.05}$\\
\hline
Strided Conv & Bilinear & ALOS-2 & $3.4\times10^{-3}$ & $24.70$ & $0.96$ & $90.42$ & $90.45$ & $80.74$ & $80.82$ \\
\hline
AvgPool & Nearest & ALOS-2 & $2.4\times10^{-2}$ & $16.28$ & $0.84$ & $82.67$ & $82.84$ & $63.14$ & $62.97$ \\
\hline
AvgPool & Bilinear & ALOS-2 & $3.9\times10^{-2}$ & $14.10$ & $0.79$ & $76.41$ & $76.81$ & $58.46$ & $57.92$ \\
\hline
\hline\hline
\textbf{Strided Conv} & \textbf{Nearest} & Bretigny & $\mathbf{1.3\times10^{-3}}$ & $\mathbf{28.71}$ & $\mathbf{0.98}$ & $\mathbf{95.74}$ & $\mathbf{95.72}$ & $\mathbf{88.98}$ & $\mathbf{88.99}$\\
\hline
Strided Conv & Bilinear & Bretigny & $4.0\times10^{-3}$ & $23.59$ & $0.94$ & $92.54$ & $92.43$ & $81.36$ & $81.47$ \\
\hline
AvgPool & Nearest & Bretigny & $4.6\times10^{-2}$ & $13.31$ & $0.68$ & $85.06$ & $84.78$ & $61.08$ & $60.66$ \\
\hline
AvgPool & Bilinear & Bretigny & $1.3\times10^{-1}$ & $8.74$ & $0.41$ & $76.49$ & $76.64$ & $47.81$ & $47.83$ \\
\hline
\end{tabular}
}
\caption{Reconstruction metrics, $H-\alpha$ classification metrics, and Cameron classification metrics are reported for the San Francisco Polarimetric SAR ALOS-2 and Brétigny datasets. Results for the ALOS-2 dataset are separated from those for Brétigny by an additional row. Across both datasets, the combination of strided convolution and nearest-neighbor layers consistently outperforms all other down-/upsampling configurations.}
\label{tab:sampling}
\end{table}
Using strided convolution for downsampling combined with nearest-neighbor upsampling yields the best and most robust performance across datasets. Compared to the strided convolution + bilinear configuration, this setup improves PSNR by approximately $5.1\,$dB on both ALOS-2 and Brétigny and reduces MSE by about $3.1\times$. In contrast, average-pooling variants perform significantly worse, with MSE values roughly 22 times higher on ALOS-2 and 100 times higher on Brétigny.
\\
\\
\textbf{Conclusion.}
Improvements in MSE, PSNR, and SSIM are consistently accompanied by gains in $H–\alpha$ and Cameron overall accuracy (OA) and F1-score, demonstrating that better preservation of amplitude and phase statistics directly enhances polarimetric semantic fidelity. Among activations, $\mathrm{CReLU}$ achieves the strongest overall performance, with $\mathrm{Cardioid}$ as a close alternative. Increasing network depth degrades performance under the current training configuration; maintaining a depth of two is optimal. Finally, the combination of strided convolution for downsampling and nearest-neighbor interpolation for upsampling best preserves polarimetric structure, while pooling and bilinear schemes should be avoided.
\subsection{Reconstruction metrics and preservation of polarimetric properties}
\label{sec:res}
\begin{figure}[htbp]
    \centering    \includegraphics[width=\columnwidth]{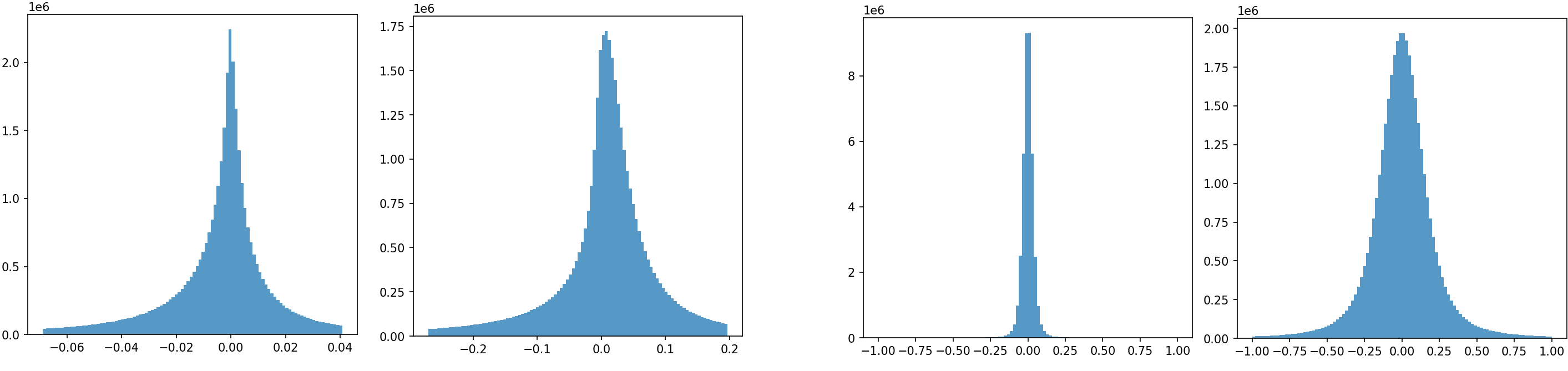}
    \caption{ Distribution of the pixel-wise reconstruction error of the amplitude (left) (with a CVNN (left) and with an RVNN (right)) and phase (right) (with a CVNN (left) and with an RVNN (right)) between the original and reconstructed images obtained on the San Francisco Polarimetric SAR ALOS-2 dataset, with reconstruction error on the $x$-axis and number of pixels on $y$-axis.}
    \label{fig:alos2_distance_analysis}
\end{figure}

\begin{figure}[htbp]
    \centering    \includegraphics[width=\columnwidth]{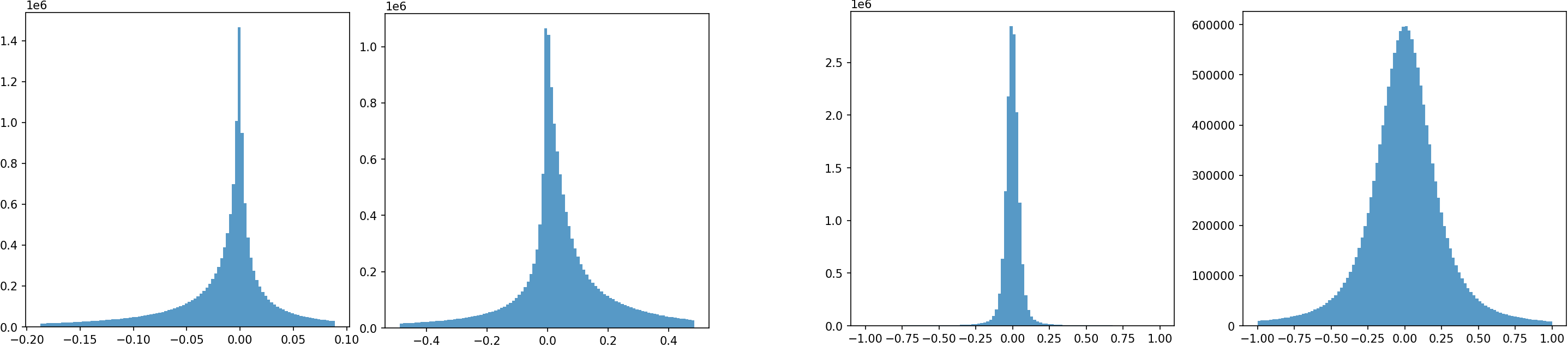}
    \caption{ Distribution of the pixel-wise reconstruction error of the amplitude (left) (with a CVNN (left) and with an RVNN (right)) and phase (right) (with a CVNN (left) and with an RVNN (right)) between the original and reconstructed images obtained on the Brétigny dataset, with reconstruction error on the $x$-axis and number of pixels on $y$-axis.}
    \label{fig:bretigny_distance_analysis}
\end{figure}

\begin{figure}[htbp]
    \centering    \includegraphics[width=\columnwidth]{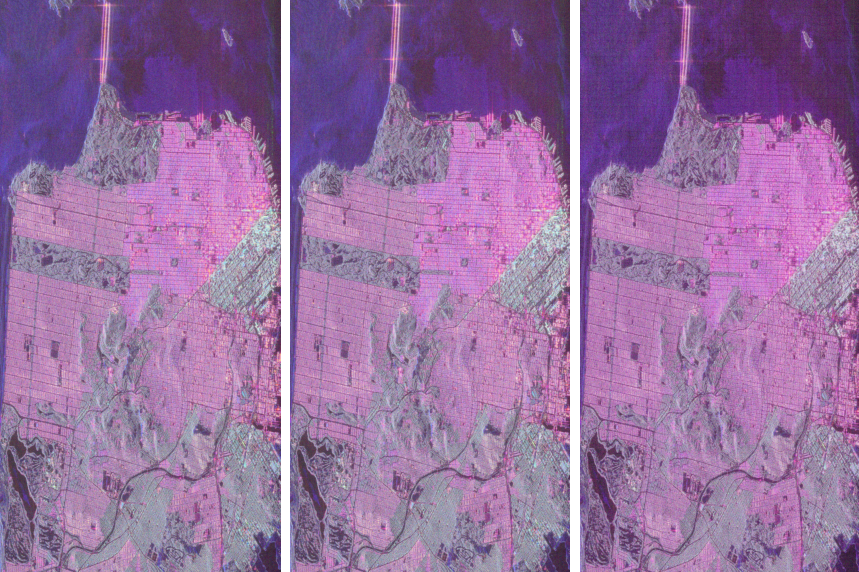}
    \caption{ Amplitude images of the original (left), reconstructed with a CVNN (middle), and reconstructed with an RVNN (right) Pauli basis images using a CVNN on the San Francisco Polarimetric SAR ALOS-2 dataset.}
    \label{fig:alos2_pauli}
\end{figure}

\begin{figure}[htbp]
    \centering    \includegraphics[width=\columnwidth]{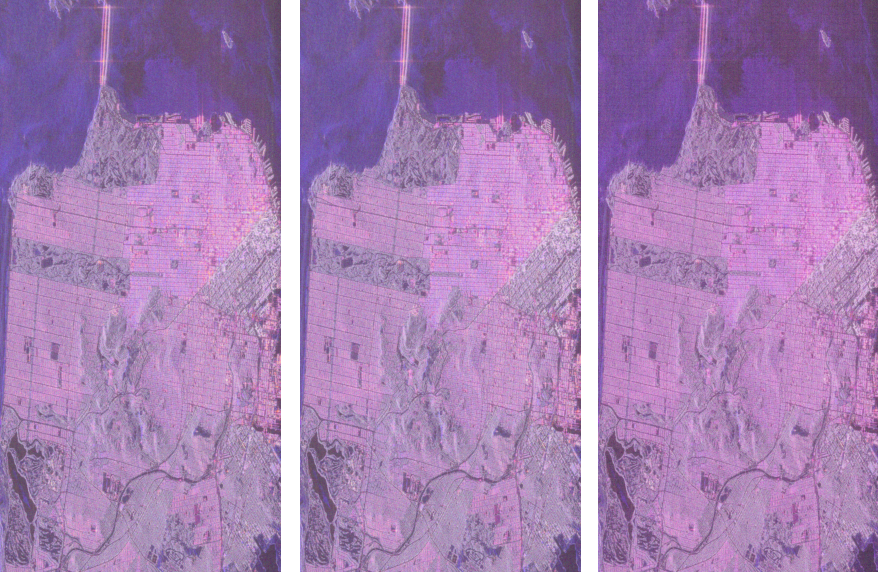}
    \caption{ Amplitude images of the original (left), reconstructed with a CVNN (middle), and reconstructed with an RVNN (right) Krogager basis images using a CVNN on the San Francisco Polarimetric SAR ALOS-2 dataset.}
    \label{fig:alos2_krogager}
\end{figure}

\begin{figure}[htbp]
    \centering    \includegraphics[width=\columnwidth]{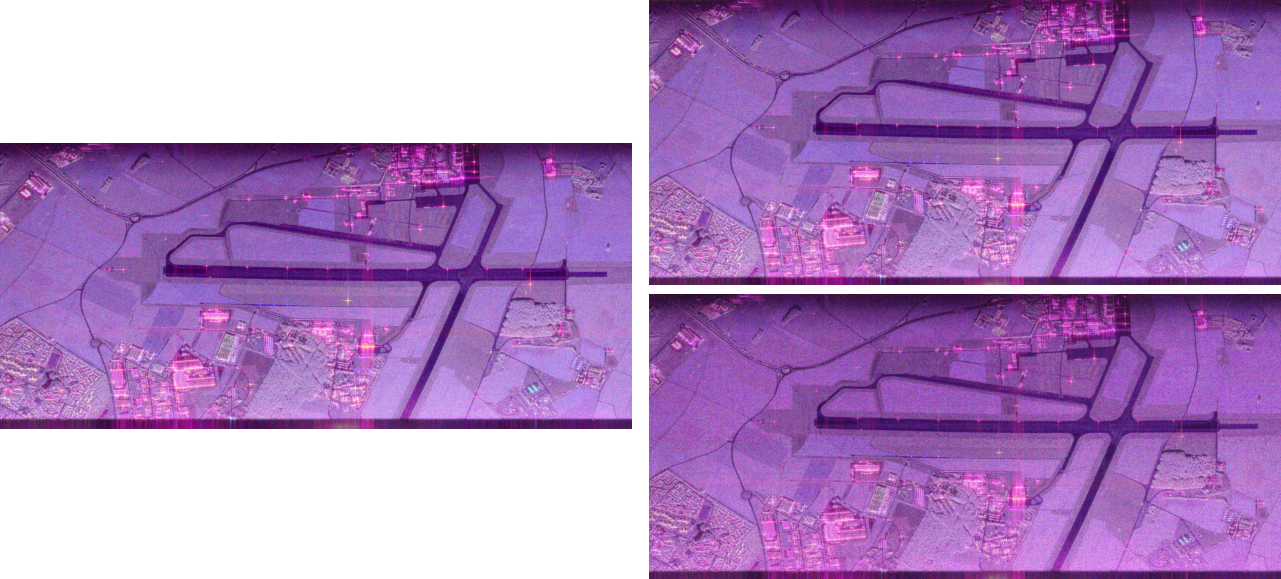}
    \caption{ Amplitude images of the original (left), reconstructed with a CVNN (up-right), and reconstructed with an RVNN (down-right) Pauli basis images using a CVNN on the Brétigny dataset.}
    \label{fig:bretigny_pauli}
\end{figure}

\begin{figure}[htbp]
    \centering    \includegraphics[width=\columnwidth]{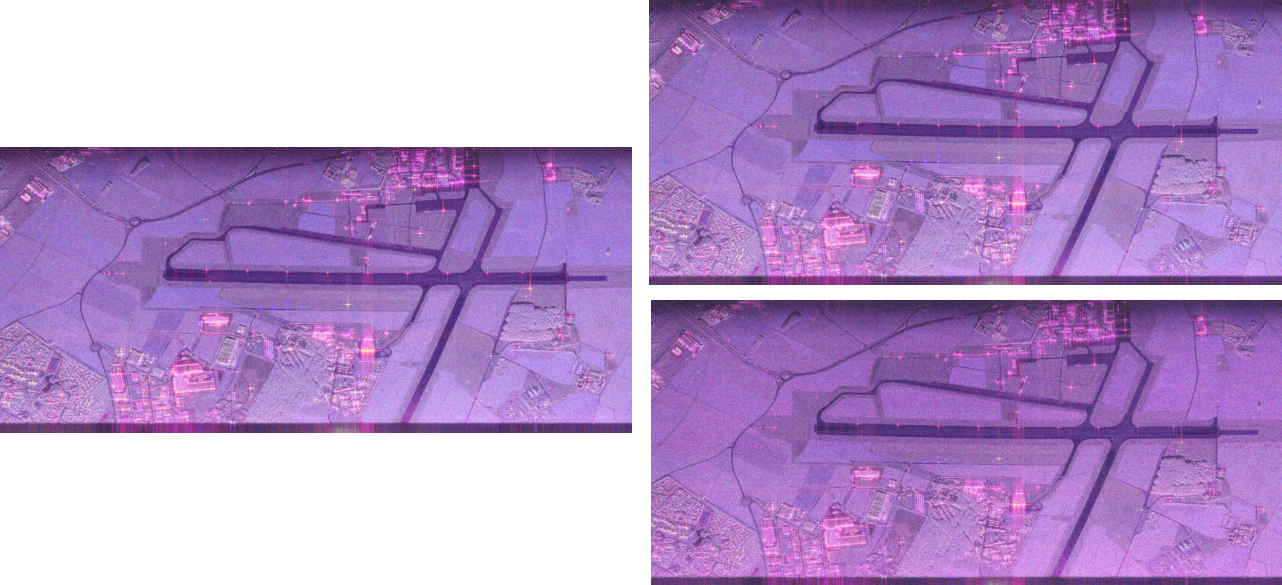}
    \caption{ Amplitude images of the original (left), reconstructed with a CVNN (up-right), and reconstructed with an RVNN (down-right) Krogager basis images using a CVNN on the Brétigny dataset.}
    \label{fig:bretigny_krogager}
\end{figure}

\begin{figure}[htbp]
    \centering    \includegraphics[width=0.95\columnwidth]{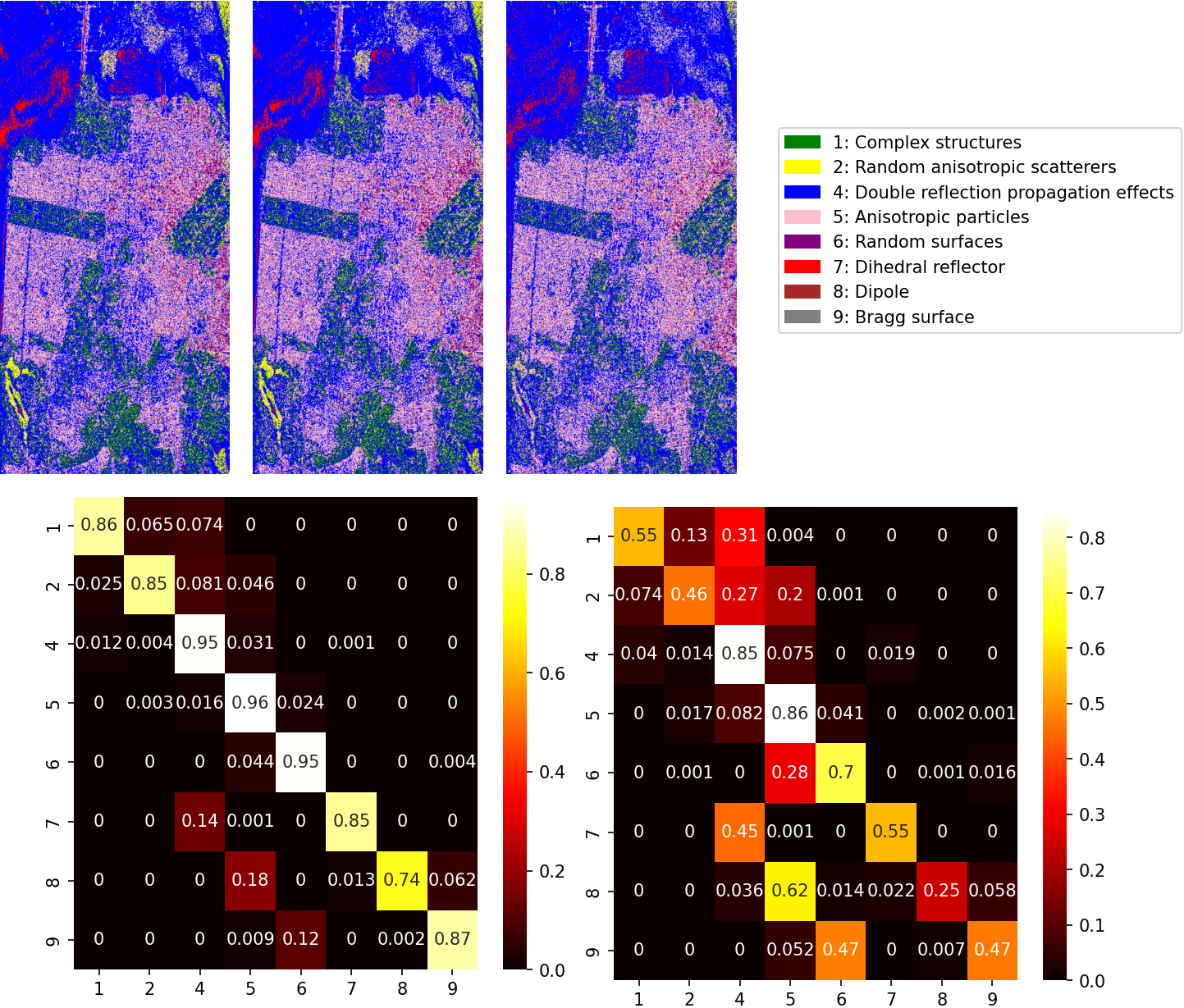}
    \caption{ Reconstruction results obtained from CVNN on the San Francisco Polarimetric SAR ALOS-2 dataset. Up: Amplitude images of the original (left), reconstructed with a CVNN (middle), and reconstructed with an RVNN (right) with the $H-\alpha$ classification. Down: Confusion matrix of the original (rows) and reconstructed (columns) $H-\alpha$ classes with a CVNN (left), and with an RVNN (right).}
    \label{fig:alos2_h_alpha}
\end{figure}

\begin{figure}[htbp]
    \centering    \includegraphics[width=0.95\columnwidth]{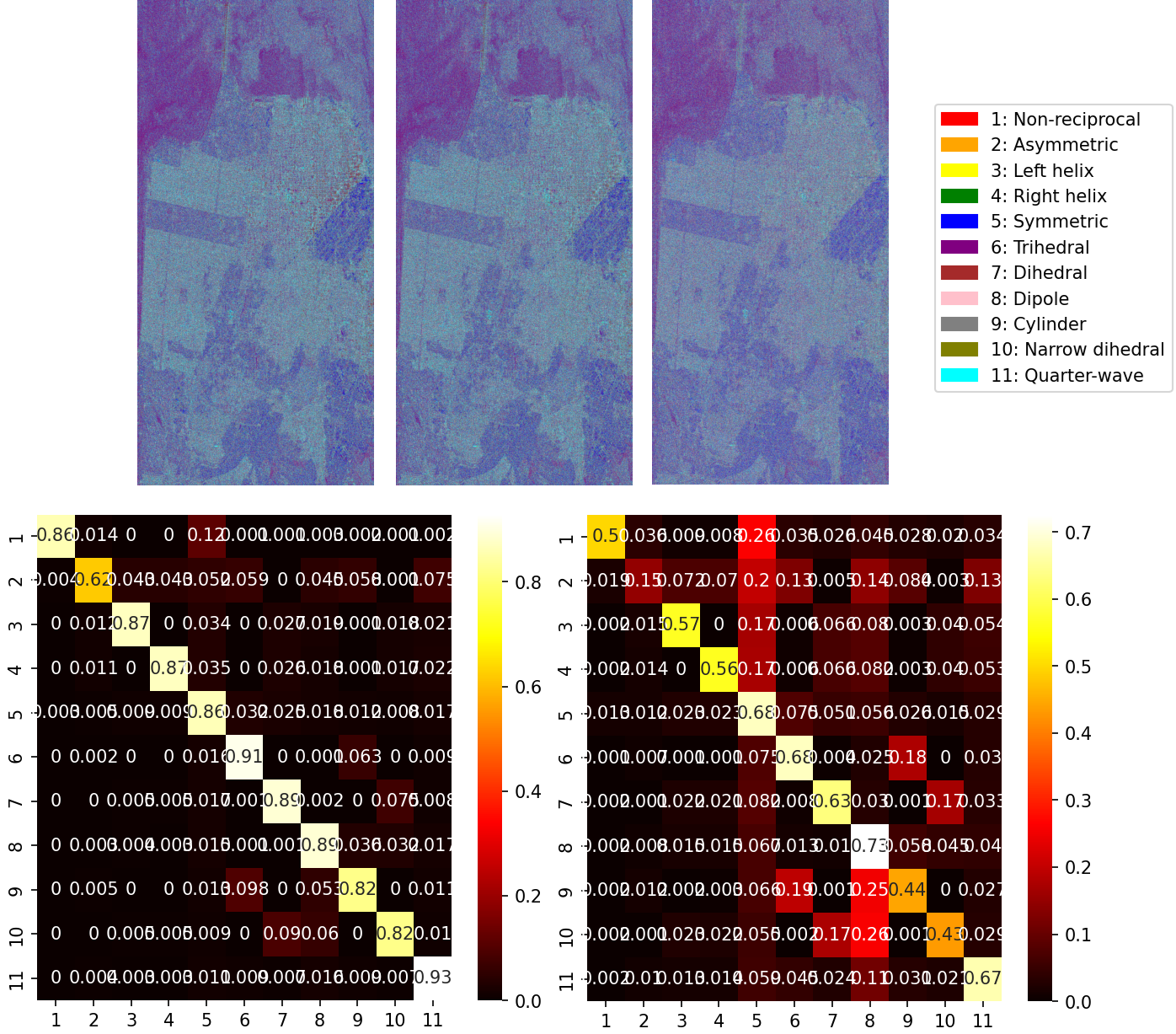}
    \caption{ Reconstruction results obtained from CVNN on the San Francisco Polarimetric SAR ALOS-2 dataset. Up: Amplitude images of the original (left), reconstructed with a CVNN (middle), and reconstructed with an RVNN (right) with the Cameron classification. Down: Confusion matrix of the original (rows) and reconstructed (columns) Cameron classes with a CVNN (left), and with an RVNN (right).}
    \label{fig:alos2_cameron}
\end{figure}

\begin{figure}[htbp]
    \centering    \includegraphics[width=0.95\columnwidth]{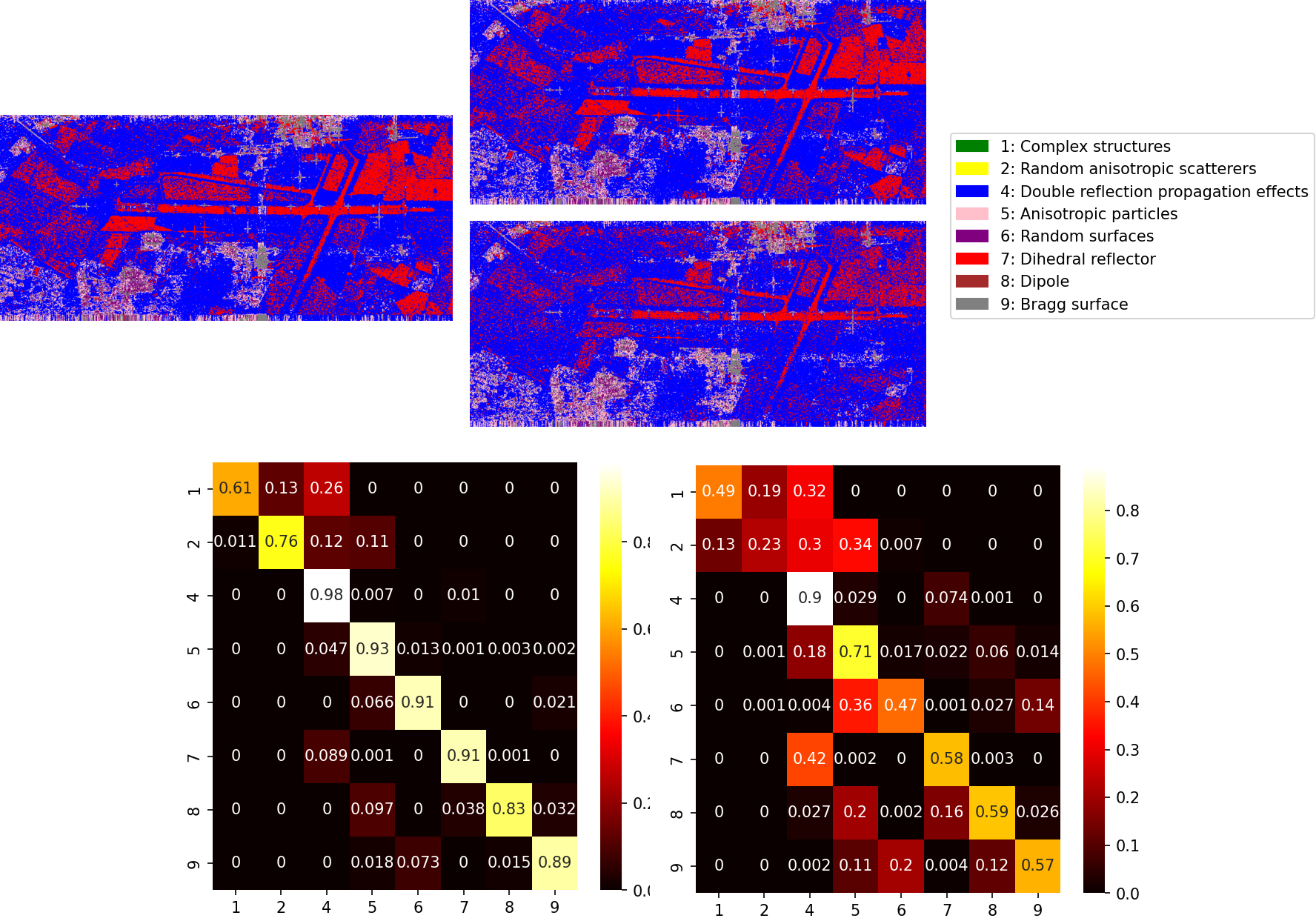}
    \caption{ Reconstruction results obtained from CVNN on the Brétigny dataset. Up: Amplitude images of the original (left), reconstructed with a CVNN (up-right), and reconstructed with an RVNN (down-right) with the $H-\alpha$ classification. Down: Confusion matrix of the original (rows) and reconstructed (columns) $H-\alpha$ classes with a CVNN (left), and with an RVNN (right).}
    \label{fig:bretigny_h_alpha}
\end{figure}

\begin{figure}[htbp]
    \centering    \includegraphics[width=0.95\columnwidth]{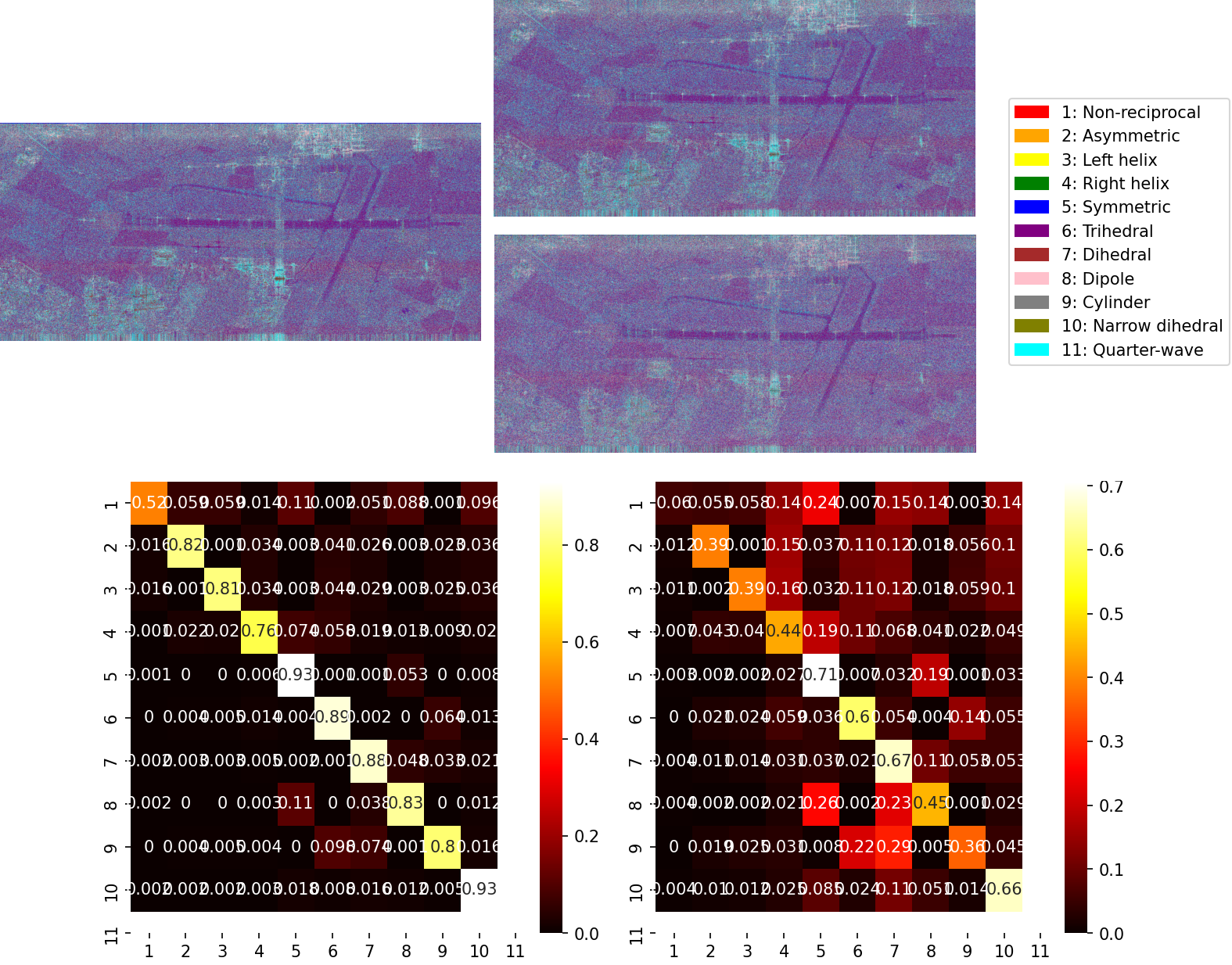}
    \caption{ Reconstruction results obtained from CVNN on the Brétigny dataset. Up: Amplitude images of the original (left), reconstructed with a CVNN (up-right), and reconstructed with an RVNN (down-right) with the Cameron classification. Down: Confusion matrix of the original (rows) and reconstructed (columns) Cameron classes with a CVNN (left), and with an RVNN (right).}
    \label{fig:bretigny_cameron}
\end{figure}

We trained the CoAE architecture (Figure~\ref{fig:coae}) using the AdamW optimizer \cite{loshchilov2017decoupled} to minimize the Mean Squared Error (MSE) loss. \cor The complete set of hyperparameters and architecture details is provided in Table~\ref{tab:hyperparameters}. \fin To ensure a fair comparison, the RVNN and CVNN models were explicitly designed to possess approximately the same number of trainable parameters. Regarding activation functions, we employed CReLU for the complex-valued models and standard ReLU for the real-valued baselines.
\begin{table}[htbp]
\centering
\resizebox{\linewidth}{!}{ 
\begin{tabular}{|c|c|c|c|c|c|c|c|c|c|c|c|}
\hline
Model & Dataset & MSE $(\downarrow)$ & PSNR $(\uparrow)$ & SSIM $(\uparrow)$ & $H$ –- $\alpha$ OA (\%) $(\uparrow)$ & $H$ –- $\alpha$ F1 (\%) $(\uparrow)$ & Cam. OA (\%) $(\uparrow)$ & Cam. F1 (\%) $(\uparrow)$ \\
\hline
\textbf{AE CVNN} & ALOS-2 & $\mathbf{1.1\times10^{-3}}$ & $\mathbf{29.77}$ & $\mathbf{0.99}$ & $\mathbf{93.81}$ & $\mathbf{93.80}$ & $\mathbf{88.03}$ & $\mathbf{88.05}$\\
\hline
AE RVNN & ALOS-2 & $2.0\times10^{-2}$ & $16.91$ & $0.83$ & $79.07$ & $78.85$ & $61.74$ & $62.20$ \\
\hline
\hline\hline
\textbf{AE CVNN} & Bretigny & $\mathbf{1.3\times10^{-3}}$ & $\mathbf{28.71}$ & $\mathbf{0.98}$ & $\mathbf{95.74}$ & $\mathbf{95.72}$ & $\mathbf{88.98}$ & $\mathbf{88.99}$ \\
\hline
AE RVNN & Bretigny & $2.6\times10^{-2}$ & $15.77$ & $0.74$ & $78.76$ & $78.31$ & $60.73$ & $61.49$ \\
\hline
\end{tabular}
}
\caption{Reconstruction metrics, $H-\alpha$ classification metrics, and Cameron classification metrics are reported for the San Francisco Polarimetric SAR ALOS-2 and Brétigny datasets. Results for the ALOS-2 dataset are separated from those for Brétigny by an additional row. Across both datasets, CVNNs consistently outperform their real-valued counterparts.}
\label{tab:cvnn_rvnn}
\end{table}
The results in Table~\ref{tab:cvnn_rvnn} support our hypothesis that CVNNs better preserve polarimetric properties. Specifically, CVNNs achieve lower MSE and higher PSNR compared to their real-valued counterparts, a trend further confirmed by the analyses presented in Figures~\ref{fig:alos2_distance_analysis}, and \ref{fig:bretigny_distance_analysis}. Indeed, we observe that both magnitude and phase errors are improved with CVNNs compared to RVNNs. This result suggests that CVNNs are more effective at pixel-wise reconstruction than their real-valued counterparts.
In addition to the quantitative metrics, Figures \ref{fig:alos2_pauli}, \ref{fig:alos2_krogager},
\ref{fig:bretigny_pauli}, and \ref{fig:bretigny_krogager} illustrate the complex-valued CoAE’s strong ability to faithfully reconstruct the original images at the pixel level, while simultaneously preserving polarimetric properties that closely match those of the ground truth.
Furthermore, the preservation of polarimetric properties is evident in the more informative 
$H-\alpha$ and Cameron decompositions, as shown in Figures \ref{fig:alos2_h_alpha}, \ref{fig:alos2_cameron}, \ref{fig:bretigny_h_alpha}, and \ref{fig:bretigny_cameron}. These representations not only demonstrate the complex-valued CoAE’s reconstruction capabilities but also highlight its limitations. The confusion matrices indicate that labels are generally well preserved during reconstruction; however, some misclassifications remain. These could potentially be mitigated by incorporating physical-awareness constraints during training or by further improving reconstruction fidelity. Figure \ref{fig:Halpha} shows that most misclassifications occur between neighboring regions of the $H-\alpha$ plane, which exhibit similar scattering properties. Since the mean squared error primarily favors the reconstruction of energy, it indirectly encourages accurate $H-\alpha$ reconstruction, but the model has no explicit awareness of the region boundaries during training. Consequently, pixels can cross class boundaries, suggesting that minimizing MSE alone may be insufficient to fully preserve more subtle polarimetric properties, such as $H-\alpha$ classification. \\

\begin{figure}[htbp]
    \centering    \includegraphics[width=\columnwidth]{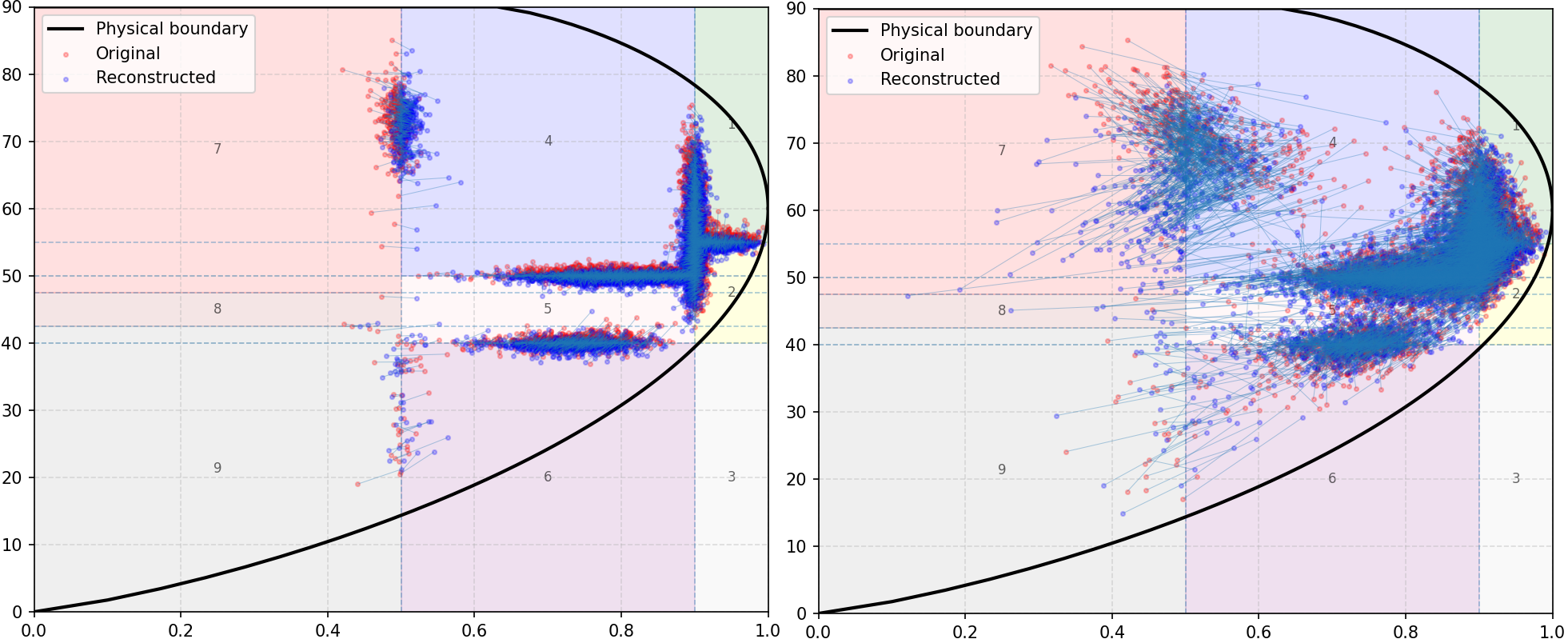}
    \caption{ Misclassification shifts on the $H-\alpha$ plane with a CVNN (left), and with an RVNN (right) on the San Francisco Polarimetric SAR ALOS-2 dataset, with the entropy on the $x$-axis and the scattering angle on the $y$-axis.}
    \label{fig:alos2_misclassification}
\end{figure}

\begin{figure}[htbp]
    \centering    \includegraphics[width=\columnwidth]{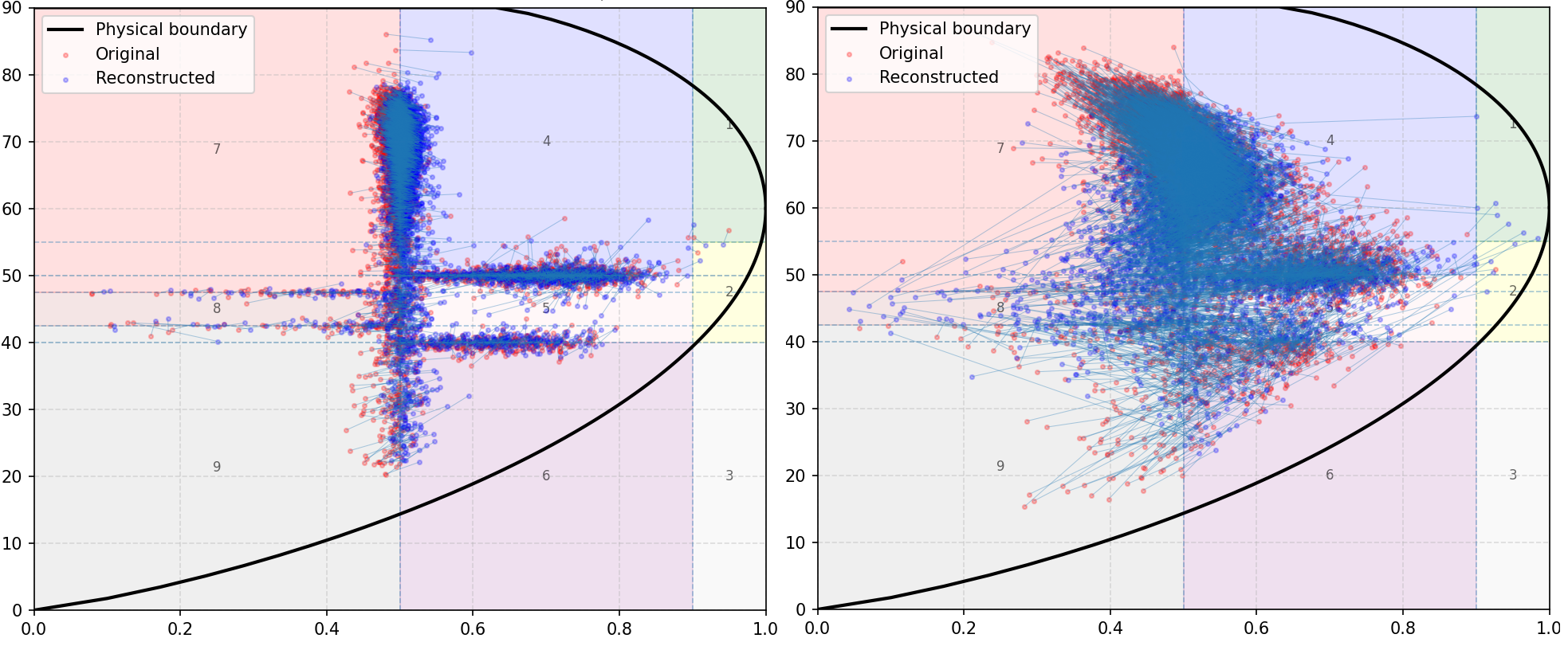}
    \caption{ Misclassification shifts on the $H-\alpha$ plane with a CVNN (left), and with an RVNN (right) on the Brétigny dataset, with the entropy on the $x$-axis and the scattering angle on the $y$-axis.}
    \label{fig:bretigny_misclassification}
\end{figure}

Finally, we analyze the misclassification shifts on the $H-\alpha$ plane illustrated by Figures \ref{fig:alos2_misclassification} and \ref{fig:bretigny_misclassification}. We observe that CVNNs suffer from fewer shifts between areas than their real-valued counterparts. Furthermore, we observe that shifts occur less frequently between non-adjacent regions when using CVNNs compared to RVNNs. This observation leads us to affirm that, in addition to better reconstruction capabilities, CVNNs induce fewer errors in polarimetric properties.

\section{Conclusion}
\label{sec:discussions}
In recent years, the combination of techniques from the field of artificial intelligence, specifically deep learning, with non-optical computer vision domains, such as SAR imaging, has become a major research topic. Indeed, due to the numerous particularities of SAR imaging, as well as the existence of advanced techniques such as polarimetric SAR, adapting standard methods from the field of deep learning computer vision is a challenging task. Additionally, most deep learning solutions found in the literature overlook the physics of the problem and apply neural networks directly to the magnitude, discarding phase information. In this work, we demonstrate the capacity of a complex-valued Convolutional \cor AutoEncoder \fin to preserve fundamental polarimetric properties throughout the reconstruction process. In other words, a latent vector representing a PolSAR image can be decoded while still retaining most of its physical properties. We have employed coherent decompositions, such as the Pauli, Krogager, and Cameron decompositions, as well as non-coherent decompositions, including the $H-\alpha$ decomposition, to validate this hypothesis. Preservation is achieved in both cases, considering that the mean squared error is the only criterion minimized during model training. This result may pave the way for more research, such as generating coherent synthetic PolSAR images.
\newpage
\bmsection*{Author Contributions}
\begin{itemize}
    \item \textbf{Quentin Gabot:} Investigation, Methodology, Software, Visualization, Writing - original draft, Writing - review \& editing.
    \item \textbf{Joana Frontera-Pons:} Supervision, Writing - original draft, Writing - review \& editing.
    \item \textbf{Jeremy Fix:} Supervision, Writing - original draft, Writing - review \& editing.
    \item \textbf{Chengfang Ren:} Supervision, Writing - original draft, Writing - review \& editing.
    \item \textbf{Jean-Philippe Ovarlez:} Supervision, Writing - original draft, Writing - review \& editing.
\end{itemize}

\bmsection*{Data Availability Statement}
\begin{itemize}
    \item \textbf{San Francisco Polarimetric SAR ALOS-2:} Publicly available at \url{https://ietr-lab.univ-rennes1.fr/polsarpro-bio/san-francisco/}.
    \item \textbf{Brétigny:} Not publicly available.
\end{itemize}

\bmsection*{Funding Information}
The authors have nothing to report.
\bmsection*{Conflict of Interest}
The authors declare no potential conflict of interest.

\bibliography{biblio}

@INPROCEEDINGS{Ramses,
  author={Dubois-Fernandez, P. and du Plessis, O.R. and le Coz, D. and Dupas, J. and Vaizan, B. and Dupuis, X. and Cantalloube, H. and Coulombeix, C. and Titin-Schnaider, C. and Dreuillet, P. and Boutry, J.-M. and Canny, J.-P. and Kaisersmertz, L. and Peyret, J. and Martineau, P. and Chanteclerc, M. and Pastore, L. and Bruyant, J.-P.},
  booktitle={IEEE International Geoscience and Remote Sensing Symposium}, 
  title={The {ONERA RAMSES SAR} system}, 
  year={2002},
  volume={3},
  pages={1723-1725}
}

@ARTICLE{Moreira,
  author={Moreira, Alberto and Prats-Iraola, Pau and Younis, Marwan and Krieger, Gerhard and Hajnsek, Irena and Papathanassiou, Konstantinos P.},
  journal={IEEE Geoscience and Remote Sensing Magazine}, 
  title={A tutorial on synthetic aperture radar}, 
  year={2013},
  volume={1},
  number={1},
  pages={6-43}
}

@article{azimi2002terrain,
  title={Terrain classification in {SAR} images using principal components analysis and neural networks},
  author={Azimi-Sadjadi, Mahmood R and Ghaloum, Saleem and Zoughi, Reza},
  journal={IEEE Transactions on Geoscience and Remote Sensing},
  volume={31},
  number={2},
  pages={511--515},
  year={2002},
  publisher={IEEE}
}

@inproceedings{dominguez2015fully,
  title={Fully polarimetric high-resolution airborne SAR image change detection with morphological component analysis},
  author={Dominguez, E Mendez and Henke, Daniel and Small, David and Meier, Erich},
  booktitle={Image and Signal Processing for Remote Sensing XXI},
  volume={9643},
  pages={374--383},
  year={2015},
  organization={SPIE}
}

@article{zhan2013sar,
  title={{SAR} image compression using multiscale dictionary learning and sparse representation},
  author={Zhan, Xin and Zhang, Rong and Yin, Dong and Huo, Chengfu},
  journal={IEEE Geoscience and Remote Sensing Letters},
  volume={10},
  number={5},
  pages={1090--1094},
  year={2013},
  publisher={IEEE}
}

@misc{loshchilov2017decoupled,
      title={Decoupled Weight Decay Regularization}, 
      author={Ilya Loshchilov and Frank Hutter},
      year={2019},
      eprint={1711.05101},
      archivePrefix={arXiv},
      primaryClass={cs.LG},
}

@article{wu2024complex,
  title={Complex-valued neurons can learn more but slower than real-valued neurons via gradient descent},
  author={Wu, Jin-Hui and Zhang, Shao-Qun and Jiang, Yuan and Zhou, Zhi-Hua},
  journal={Advances in Neural Information Processing Systems},
  volume={36},
  year={2024}
}

@article{li2008complex,
  title={Complex-valued adaptive signal processing using nonlinear functions},
  author={Li, Hualiang and Adal{\i}, T{\"u}lay},
  journal={EURASIP Journal on Advances in Signal Processing},
  volume={},
  pages={1--9},
  year={2008},
  publisher={Springer}
}

@article{trabelsi2017deep,
  title={Deep complex networks},
  author={Trabelsi, Chiheb and Bilaniuk, Olexa and Zhang, Ying and Serdyuk, Dmitriy and Subramanian, Sandeep and Santos, Joao Felipe and Mehri, Soroush and Rostamzadeh, Negar and Bengio, Yoshua and Pal, Christopher J},
  journal={arXiv preprint arXiv:1705.09792},
  year={2017}
}

@article{barrachina2023comparison,
  title={Comparison between equivalent architectures of complex-valued and real-valued neural networks-Application on polarimetric {SAR} image segmentation},
  author={Barrachina, J. A. and Ren, Chengfang and Morisseau, Christ{\`e}le and Vieillard, Gilles and Ovarlez, J.-P.},
  journal={Journal of Signal Processing Systems},
  volume={95},
  number={1},
  pages={57--66},
  year={2023},
  publisher={Springer}
}

@INPROCEEDINGS{gabotpreserving,
  author={Gabot, Quentin and Fix, Jérémy and Frontera-Pons, Joana and Ren, Chengfang and Ovarlez, Jean-Philippe},
  booktitle={2024 International Radar Conference (RADAR)}, 
  title={Preserving polarimetric properties in {PolSAR} image reconstruction through Complex-Valued Auto-Encoders}, 
  year={2024},
  volume={},
  number={},
  pages={}
}

@article{zhu2021deep,
  title={Deep {L}earning meets {SAR}: Concepts, models, pitfalls, and perspectives},
  author={Zhu, X. X. and Montazeri, Sina and Ali, Mohsin and Hua, Yuansheng and Wang, Yuanyuan and Mou, Lichao and Shi, Yilei and Xu, Feng and Bamler, Richard},
  journal={IEEE Geoscience and Remote Sensing Magazine},
  volume={9},
  number={4},
  pages={143--172},
  year={2021},
  publisher={IEEE}
}

@article{cole2021analysis,
  title={Analysis of deep complex-valued convolutional neural networks for {MRI} reconstruction and phase-focused applications},
  author={Cole, Elizabeth and Cheng, Joseph and Pauly, John and Vasanawala, Shreyas},
  journal={Magnetic resonance in medicine},
  volume={86},
  number={2},
  pages={1093--1109},
  year={2021},
  publisher={Wiley Online Library}
}

@INPROCEEDINGS{Barrachina2021,
  author={Barrachina, J. A.  and Ren, Chengfang and Morisseau, Christ{\`e}le and Vieillard, Gilles and Ovarlez, J.-P.},
  booktitle={IEEE International Conference on Acoustics, Speech and Signal Processing (ICASSP)}, 
  title={Complex-Valued Vs. Real-Valued Neural Networks for Classification Perspectives: An Example on Non-Circular Data}, 
  year={2021},
  volume={},
  number={},
  pages={2990-2994}}

@article{lee2022complex,
  title={Complex-valued neural networks: A comprehensive survey},
  author={Lee, ChiYan and Hasegawa, Hideyuki and Gao, Shangce},
  journal={IEEE/CAA Journal of Automatica Sinica},
  volume={9},
  number={8},
  pages={1406--1426},
  year={2022},
  publisher={IEEE}
}

@inproceedings{kuroe2003activation,
  title={On activation functions for complex-valued neural networks—existence of energy functions},
  author={Kuroe, Yasuaki and Yoshid, Mitsuo and Mori, Takehiro},
  booktitle={International Conference on Artificial Neural Networks},
  pages={985--992},
  year={2003},
  organization={Springer}
}

@inproceedings{arjovsky2016unitary,
  title={Unitary evolution recurrent neural networks},
  author={Arjovsky, Martin and Shah, Amar and Bengio, Yoshua},
  booktitle={International conference on machine learning},
  pages={1120--1128},
  year={2016},
  organization={PMLR}
}

@article{sarroff2015learning,
  title={Learning representations using complex-valued nets},
  author={Sarroff, Andy M. and Shepardson, Victor and Casey, Michael A.},
  journal={arXiv preprint arXiv:1511.06351},
  year={2015}
}

@INPROCEEDINGS{yamaguchi2016alos,
  author={Yamaguchi, Y. and Sato, R. and Yamada, H.},
  booktitle={Proceedings of EUSAR 2016: 11th European Conference on Synthetic Aperture Radar}, 
  title={{ALOS-2} Quad. Pol. Images and {ALOS} Ones}, 
  year={2016},
  volume={},
  number={},
  pages={1-4}}

@inproceedings{he2016deep,
  title={Deep residual learning for image recognition},
  author={He, Kaiming and Zhang, Xiangyu and Ren, Shaoqing and Sun, Jian},
  booktitle={Proceedings of the IEEE conference on computer vision and pattern recognition},
  pages={770--778},
  year={2016}
}

@inproceedings{ronneberger2015u,
  title={U-net: Convolutional networks for biomedical image segmentation},
  author={Ronneberger, Olaf and Fischer, Philipp and Brox, Thomas},
  booktitle={Medical image computing and computer-assisted intervention--MICCAI 2015: 18th international conference, Munich, Germany, October 5-9, 2015, proceedings, part III 18},
  pages={234--241},
  year={2015},
  organization={Springer}
}

@article{hinton1993autoencoders,
  title={Autoencoders, minimum description length and {H}elmholtz free energy},
  author={Hinton, G. E. and Zemel, Richard},
  journal={Advances in neural information processing systems},
  volume={6},
  year={1993}
}

@article{mian2019design,
  title={Design of new wavelet packets adapted to high-resolution {SAR} images with an application to target detection},
  author={Mian, Ammar and Ovarlez, J.-P. and Atto, A. M. and Ginolhac, Guillaume},
  journal={IEEE Transactions on Geoscience and Remote Sensing},
  volume={57},
  number={6},
  pages={3919--3932},
  year={2019},
  publisher={IEEE}
}

@inproceedings{dedmari2018complex,
  title={Complex fully convolutional neural networks for {MR} image reconstruction},
  author={Dedmari, M. A. and Conjeti, Sailesh and Estrada, Santiago and Ehses, Phillip and St{\"o}cker, Tony and Reuter, Martin},
  booktitle={International Workshop on Machine Learning for Medical Image Reconstruction},
  pages={30--38},
  year={2018},
  organization={Springer}
}

@article{patruno2013polarimetric,
  title={Polarimetric Multifrequency and Multi-incidence {SAR} Sensors Analysis for Archaeological Purposes},
  author={Patruno, Jolanda and Dore, Nicole and Crespi, Mattia and Pottier, Eric},
  journal={Archaeological Prospection},
  volume={20},
  number={2},
  pages={89--96},
  year={2013},
  publisher={Wiley Online Library}
}

@book{quegan_understanding_2004,
  title={Understanding {S}ynthetic {A}perture {R}adar Images},
  author={Oliver, C. and Quegan, S.},
  isbn={9781891121319},
  series={SciTech radar and defense series},
  year={2004},
  publisher={SciTech Publishing, Inc}
}

@article{formont2010statistical,
  title={Statistical classification for heterogeneous polarimetric {SAR} images},
  author={Formont, P. and Pascal, F. and Vasile, G. and Ovarlez, J.-P. and Ferro-Famil, L.},
  journal={IEEE Journal of Selected Topics in Signal Processing},
  volume={5},
  number={3},
  pages={567--576},
  year={2010},
  publisher={IEEE}
}

@book{henderson1998manual,
  title={Manual of remote sensing: principles and applications of imaging radar},
  author={Henderson, F. M. and Lewis, A. J.},
  year={1998},
  publisher={Wiley}
}

@article{KrogagerTGRS1993,
  author  = {E. Krogager},
  title   = {A New Decomposition of the Radar Target Scattering Matrix},
  journal = {IEEE Transactions on Geoscience and Remote Sensing},
  year    = {1993},
  volume  = {31},
  number  = {4},
  pages   = {},
  note    = {Krogager's coherent triad model (sphere--diplane--helix)},
}

@inproceedings{Krogager92,
 author        = "E. Krogager",
 title         = "Utilization and Interpretation of Polarimetric Data in High Resolution Radar Target Imaging",
 booktitle     = "Proc. Second International Workshop on Radar Polarimetry ({JIPR}'1992)",
 volume        = "2",
 address       = "Nantes, France",
 month         = sep # " 8--10,",
 year          = "1992",
 pages         = "547-557"
}

@inproceedings{Krogager95,
 author        = "E. Krogager and Z. H. Czyz",
 title         = "Properties of the Sphere, Diplane, Helix Decomposition",
 booktitle     = "Proc. Third International Workshop on Radar Polarimetry ({JIPR}'1995)",
 volume        = "1",
 address       = "Nantes, France",
 month         = mar # " 21--23,",
 year          = "1995",
 pages         = "106-114"
}

@inproceedings{Krogager95A,
 author        = "E. Krogager and J. Dall and S. N. Madsen",
 title         = "The Sphere, Diplane, Helix Decomposition Recent Results With Polarimetric {SAR} Data",
 booktitle     = "Proc. Third International Workshop on Radar Polarimetry ({JIPR}'1995)",
 volume        = "2",
 address       = "Nantes, France",
 month         = mar # " 21--23,",
 year          = "1995",
 pages         = "621-625"
}

@misc{lowe_complex-valued_2022,
    title = {Complex-{Valued} {Autoencoders} for {Object} {Discovery}},
    language = {en},
    publisher = {arXiv},
    author = {Löwe, Sindy and Lippe, Phillip and Rudolph, Maja and Welling, Max},
    month = nov,
    year = {2022},
    note = {arXiv:2204.02075 [cs]},
}

@article{henderson1997sar,
  title={{SAR} applications in human settlement detection, population estimation and urban land use pattern analysis: a status report},
  author={Henderson, F. M. and Xia, Z.-G.},
  journal={IEEE Transactions on Geoscience and Remote Sensing},
  volume={35},
  number={1},
  pages={79--85},
  year={1997},
  publisher={IEEE}
}

@inproceedings{frontera2023unsupervised,
  title={Unsupervised {SAR} change detection with despeckling autoencoders},
  author={Frontera-Pons, J. and Brigui, F. and De Milly, X.},
  booktitle={2023 IEEE International Radar Conference (RADAR)},
  pages={1--6},
  year={2023},
  organization={IEEE}
}

@article{shang_complex-valued_2019,
    title = {Complex-{Valued} {Convolutional} {Autoencoder} and {Spatial} {Pixel}-{Squares} {Refinement} for {Polarimetric} {SAR} {Image} {Classification}},
    volume = {11},
    issn = {2072-4292},
    language = {en},
    number = {5},
    journal = {Remote Sensing},
    author = {Shang, Ronghua and Wang, Guangguang and A. Okoth, Michael and Jiao, Licheng},
    month = mar,
    year = {2019},
    pages = {522},
}

@article{rumelhart_learning_1986,
	title = {Learning representations by back-propagating errors},
	volume = {323},
	issn = {0028-0836, 1476-4687},
	language = {en},
	number = {6088},
	journal = {Nature},
	author = {Rumelhart, D. E. and Hinton, G. E. and Williams, R. J.},
	month = oct,
	year = {1986},
	pages = {533--536},
}

@article{baldi2012complex,
  title={Complex-valued autoencoders},
  author={Baldi, P. and Lu, Z.},
  journal={Neural Networks},
  volume={33},
  pages={136--147},
  year={2012},
  publisher={Elsevier}
}

@incollection{honkela_stacked_2011,
	address = {Berlin, Heidelberg},
	title = {Stacked {Convolutional} {Auto}-{Encoders} for {Hierarchical} {Feature} {Extraction}},
	volume = {6791},
	isbn = {978-3-642-21734-0 978-3-642-21735-7},
	booktitle = {Artificial {Neural} {Networks} and {Machine} {Learning} – {ICANN} 2011},
	publisher = {Springer Berlin Heidelberg},
	author = {Masci, J. and Meier, U. and Cireşan, D. and Schmidhuber, J.},
	editor = {Honkela, T. and Duch, W. and Girolami, M. and Kaski, S.},
	year = {2011},
	note = {Series Title: Lecture Notes in Computer Science},
	pages = {52--59},
}

@inproceedings{asiyabi2022complex,
  title={Complex-valued autoencoders with coherence preservation for {SAR}},
  author={Asiyabi, Reza Mohammadi and Datcu, Mihai and Anghel, Andrei and Nies, Holger},
  booktitle={EUSAR 2022; 14th European Conference on Synthetic Aperture Radar},
  pages={1--6},
  year={2022},
  organization={VDE}
}

@inproceedings{cameron1990feature,
  title={Feature motivated polarization scattering matrix decomposition},
  author={Cameron, William L and Leung, Ling K},
  booktitle={IEEE International Conference on Radar},
  pages={549--557},
  year={1990},
  organization={IEEE}
}

@article{cameron2002simulated,
  title={Simulated polarimetric signatures of primitive geometrical shapes},
  author={Cameron, William L and Youssef, Nazih N and Leung, Ling K},
  journal={IEEE Transactions on Geoscience and Remote Sensing},
  volume={34},
  number={3},
  pages={793--803},
  year={2002},
  publisher={IEEE}
}

@article{cameron2006conservative,
  title={Conservative polarimetric scatterers and their role in incorrect extensions of the {C}ameron decomposition},
  author={Cameron, William L. and Rais, Houra},
  journal={IEEE Transactions on Geoscience and Remote Sensing},
  volume={44},
  number={12},
  pages={3506--3516},
  year={2006},
  publisher={IEEE}
}

@article{cloude1996review,
  title={A review of target decomposition theorems in radar polarimetry},
  author={Cloude, Shane R. and Pottier, Eric},
  journal={IEEE Transactions on Geoscience and Remote Sensing},
  volume={34},
  number={2},
  pages={498--518},
  year={1996},
  publisher={IEEE}
}

@book{lee2017polarimetric,
  title={Polarimetric radar imaging: from basics to applications},
  author={Lee, Jong-Sen and Pottier, Eric},
  year={2017},
  publisher={CRC press}
}

@article{cloude2002entropy,
  title={An entropy based classification scheme for land applications of polarimetric {SAR}},
  author={Cloude, Shane R. and Pottier, Eric},
  journal={IEEE Transactions on Geoscience and Remote Sensing},
  volume={35},
  number={1},
  pages={68--78},
  year={2002},
  publisher={IEEE}
}

@inproceedings{alberga2004potential,
  title={Potential of coherent decompositions in {SAR} polarimetry and interferometry},
  author={Alberga, Vito and Krogager, Ernst and Chandra, Madhukar and Wanielik, Gerd},
  booktitle={IGARSS 2004. 2004 IEEE International Geoscience and Remote Sensing Symposium},
  volume={3},
  pages={1792--1795},
  year={2004},
  organization={IEEE}
}

@article{pauli1941relativistic,
  title={Relativistic field theories of elementary particles},
  author={Pauli, Wolfgang},
  journal={Reviews of Modern Physics},
  volume={13},
  number={3},
  pages={203},
  year={1941},
  publisher={APS}
}

@article{freeman2002three,
  title={A three-component scattering model for polarimetric {SAR} data},
  author={Freeman, Anthony and Durden, Stephen L},
  journal={IEEE Transactions on Geoscience and Remote Sensing},
  volume={36},
  number={3},
  pages={963--973},
  year={2002},
  publisher={IEEE}
}

@article{yamaguchi2005four,
  title={Four-component scattering model for polarimetric {SAR} image decomposition},
  author={Yamaguchi, Yoshio and Moriyama, Toshifumi and Ishido, Motoi and Yamada, Hiroyoshi},
  journal={IEEE Transactions on Geoscience and Remote Sensing},
  volume={43},
  number={8},
  pages={1699--1706},
  year={2005},
  publisher={IEEE}
}

@article{kingma_auto-encoding_2013,
    title = {Auto-{Encoding} {Variational} {Bayes}},
    copyright = {arXiv.org perpetual, non-exclusive license},
    author = {Kingma, Diederik P and Welling, Max},
    year = {2013},
    note = {Publisher: [object Object]
Version Number: 11},
}

@article{Valle2024,
title = {Universal approximation theorem for vector- and hypercomplex-valued neural networks},
journal = {Neural Networks},
volume = {180},
pages = {106632},
year = {2024},
issn = {0893-6080},
author = {Marcos Eduardo Valle and Wington L. Vital and Guilherme Vieira}
}

@INPROCEEDINGS{Eilers2023,
  author={Eilers, Florian and Jiang, Xiaoyi},
  booktitle={IEEE International Conference on Acoustics, Speech and Signal Processing (ICASSP)}, 
  title={Building Blocks for a Complex-Valued Transformer Architecture}, 
  year={2023},
  volume={},
  number={},
  pages={1-5}
}

@article{glorot_understanding_2010,
    title = {Understanding the difﬁculty of training deep feedforward neural networks},
    language = {en},
    author = {Glorot, Xavier and Bengio, Yoshua},
    year = {2010},
}

@inproceedings{he_delving_2015,
	location = {Santiago, Chile},
	title = {Delving Deep into Rectifiers: Surpassing Human-Level Performance on {ImageNet} Classification},
	isbn = {978-1-4673-8391-2},
	shorttitle = {Delving Deep into Rectifiers},
	eventtitle = {2015 {IEEE} International Conference on Computer Vision ({ICCV})},
	pages = {1026--1034},
	booktitle = {2015 {IEEE} International Conference on Computer Vision ({ICCV})},
	publisher = {{IEEE}},
	author = {He, Kaiming and Zhang, Xiangyu and Ren, Shaoqing and Sun, Jian},
	date = {2015},
}

@inproceedings{Ioffe2015,
author = {Ioffe, Sergey and Szegedy, Christian},
title = {Batch normalization: accelerating deep network training by reducing internal covariate shift},
year = {2015},
publisher = {JMLR.org},
booktitle = {Proceedings of the 32nd International Conference on International Conference on Machine Learning - Volume 37},
pages = {448–456},
numpages = {9},
location = {Lille, France},
series = {ICML'15}
}

@inproceedings{Zhang2019,
    address = "Vancouver, Canada",
    author = "Zhang, Biao and Sennrich, Rico",
    booktitle = "Advances in Neural Information Processing Systems 32",
    title = "{Root Mean Square Layer Normalization}",
    year = "2019"
}

@misc{Ba2016,
      title={Layer Normalization}, 
      author={Jimmy Lei Ba and Jamie Ryan Kiros and Geoffrey E. Hinton},
      year={2016},
      eprint={1607.06450},
      archivePrefix={arXiv},
      primaryClass={stat.ML},
}

@article{Odena2016,
  author = {Odena, Augustus and Dumoulin, Vincent and Olah, Chris},
  title = {Deconvolution and Checkerboard Artifacts},
  journal = {Distill},
  year = {2016},
}

@article{barrachina2023theory,
  title={Theory and implementation of complex-valued neural networks},
  author={Barrachina, Jose Agustin and Ren, Chengfang and Vieillard, Gilles and Morisseau, Christele and Ovarlez, Jean-Philippe},
  journal={arXiv preprint arXiv:2302.08286},
  year={2023}
}

@article{guberman2016complex,
  title={On complex valued convolutional neural networks},
  author={Guberman, Nitzan},
  journal={arXiv preprint arXiv:1602.09046},
  year={2016}
}

@article{gao2018enhanced,
  title={Enhanced radar imaging using a complex-valued convolutional neural network},
  author={Gao, Jingkun and Deng, Bin and Qin, Yuliang and Wang, Hongqiang and Li, Xiang},
  journal={IEEE Geoscience and Remote Sensing Letters},
  volume={16},
  number={1},
  pages={35--39},
  year={2018},
  publisher={IEEE}
}

@inproceedings{virtue2017better,
  title={Better than real: Complex-valued neural nets for {MRI} fingerprinting},
  author={Virtue, Patrick and Stella, X Yu and Lustig, Michael},
  booktitle={2017 IEEE international conference on image processing (ICIP)},
  pages={3953--3957},
  year={2017},
  organization={IEEE}
}

\end{document}